# Architectures for Protecting Cloud Data Planes

Grant Dasher, Google Inc.
Ines Envid, Google Inc.
Brad Calder, Google Inc.
Jan 2022

**Abstract**

This paper explores three approaches for protecting cloud application data planes to prevent unauthorized access to the application and its data and to prevent unwanted data exfiltration. Through an exploration of various concrete security architectures, we focus on (1) Cloud Security Perimeters to provide a boundary around data and infrastructure in the cloud that provides a line of defense both to improper access to sensitive information and the exfiltration of that information, (2) Cloud Landing Points to provide a safe integration point between parts of your cloud applications and on-premises applications to communicate through, and (3) Zero Trust security architectures that are built on the principles of defense in depth and least-privilege access. Using these approaches together provides critical protection for services and applications as they transition from traditional on-premises network security to the Cloud security architectures, and then to potentially Zero Trust security architectures.

**Section 1: Introduction**

Protecting services and applications is a critically important part of cybersecurity. National Institute of Standards and Technology (NIST) divides the overall cybersecurity problem space into 5 categories: "Identify", "Protect", "Detect", "Respond", and "Recover"[1]. In this paper, we will focus on the "Protect" category (and to a lesser degree "Detect") for cloud-hosted applications, with a bias towards the network and identity based technologies that are commonly used in such an environment. We acknowledge that this is not a comprehensive treatment of cloud security, but it is an important part of protecting applications, and often at the root of real-world incidents (e.g., incidents of exposed and unsecured data stored in the cloud). There are many other areas of security that are beyond the scope of this paper, such as OS security, encryption, hardware-level security, and software supply chain security.

Cloud computing has pushed traditional network-centric approaches to security to their limits. The proliferation of devices, compute types, SaaS/PaaS services, and public cloud has challenged traditional security concepts and opened the doors to new models such as Zero Trust. Zero Trust is a broad term capturing a set of related ideas around the concept that applications should not trust the network and should authenticate all attempts to access their data [2]. However, articulating a concrete path to transition from traditional network security to Zero Trust requires architectures in which both different models and solutions coexist and work together. Our paper examines various architectures for securing cloud application data planes and explores how to transition from traditional on-premises network security architectures, to Cloud security architectures, and then how they can be combined with architectures compatible with Zero Trust security principles.

From a theoretical point of view, within the space of cloud networking and identity, one typically seeks various protection goals. These are usually defined in response to an explicit *threat model* which outlines the problems one is concerned with and the assumptions one makes about one's adversaries. In this paper, we will assume a relatively sophisticated adversary and that there are three high priority goals one is seeking to defend against:

1. **Prevent Unauthorized Access** - Control and restrict the access and operation privileges in applications, infrastructure and workloads, to their authorized client humans and/or client applications, from authorized locations. This includes preventing unauthorized software (malware) from entering the environment as it poses a vector for unauthorized access to data and data exfiltration.
2. **Prevent Exfiltration** - A client human or client application is authorized to access the data within a given scope, but the data cannot be taken to any location of the client's choice. The data must stay in authorized locations and subject to intended controls that protect against possible accidental misconfigurations of cloud applications (e.g., a storage bucket resource, or against malicious access to the application).
3. **Prevent Lateral Movement** - When a compromise occurs in an application, limit the ability for the intruder to move to other applications or workloads.

We will build out a conceptual model for thinking about cloud architectures in the context of security. We will assume that applications and workloads consist of sets of services that can be secured. We will then leverage this conceptual model to define various architectures for achieving the above security goals. As we will do this, we will look at:

- **Communication within an Application** - It is important to protect data as services within an application talk to each other.
- **Application Inbound Communication** - It is important to provide secure and granular access to an application's services, its data and the cloud provider's managed services, when being accessed from the Internet, from on-premises or from services hosted on the same or other cloud platform.
- **Application Outbound Communication** - It is important to filter and secure the application's data when the application can connect to internet-based services, on-premises services, or services hosted on the same, or other, cloud platforms.

We will also examine how the organizational structure of application ownership influences the application of security policies. These policies then translate into specific network topologies as one of the ways of enforcing policy goals. Thus the network topologies often align to organizational boundaries. We examine how to build an application, secure by default, with cloud providers providing adequate infrastructure to apply security policies (authentication, authorization and auditing), and enforcement points to protect the application from untrusted workloads and malicious actors. Additionally, we will describe, step-by-step, how to enhance the security posture of an existing application.

In each of these protection architectures, there is a requirement to monitor and audit the compliance and state of the applications and to provide visibility into the policy enforcement for debugging. The details on how to achieve this are not the primary focus of this paper, although we will provide some high level guidance when comparing the architectural models in Section 4.

**Section 2: Core Concepts for Protection**

Cloud is an extremely broad term, spanning a variety of technologies ranging from Infrastructure-as-a-Service (IaaS) technologies like VMs and virtual networks, through Platform-as-a-Service (PaaS) technologies like AWS Lambda, Cloud Run in GCP, and managed data products such as BigQuery, to Software-as-a-Service (SaaS) products.

Historically, in on-premises networks, the predominant approach to security was to build a *perimeter* around one's network to keep adversaries off the network, and to assume that all entities on the network are relatively trustworthy. This perimeter abstraction is a logical concept that was implemented primarily with network-centric controls, like firewalls and security gateway devices. On-premises, the control/management plane (e.g., for configuring vSphere-based environments) was typically collocated within the data plane (e.g. the infrastructure/services enabling the transfer of data), resulting in network perimeters enclosing both the control/management and the data planes. In the on-premises world, networks could even be physically separated (e.g., with an air gap), which is not typically possible in the cloud model. In the cloud, tools like access policies and API governance must be used to secure the control/management plane. The topic of cloud control plane security is vitally important, but not the focus of this paper.

The network security perimeter used on-premises remains relevant in the cloud as a security boundary, but it is in need of modernization to account for cloud technologies. For example, in the cloud the application owner has limited control of the underlying infrastructure stack (e.g., the IP addresses to the compute servers), where both the control plane and many of the services are cloud-native and multi-tenant. This means the concept of a perimeter, especially a network-centric perimeter, applies very differently in the cloud. We will discuss how to incorporate a perimeter into a cloud security architecture in Section 3.2.1.

Another concept we focus on in this paper is what we call a **Cloud Landing Point** (CLP). A CLP provides isolation of individual cloud workloads from each other and from the legacy corporate network. Different classes of workloads can be protected by their own cloud security perimeters and then combined together with CLP. These two key concepts (security perimeters and landing points) help define an architecture for the cloud data plane--that is, how application clients connect to workloads and how workloads talk to each other.

- **Cloud Security Perimeters** - provide a boundary around data and infrastructure in the cloud that provides a line of defense both to improper access to sensitive information and the exfiltration of that information outside the enterprise. Although perimeters can be implemented in the cloud in a similar way to their usage on-premises, they can also be extended to use new types of cloud native primitives like the cloud resource hierarchy.
- **Cloud Landing Points** - provides a safe environment for workload deployment built around chokepoints for enforcing uniform policies like firewalls for traffic control, identity authentication/authorization, and traffic inspection across various types of workloads. Our focus in this paper is on the data plane.

We describe the above concepts and how they are applied to services and users within an organization. For protecting against the sharing of data from a given organization's user when located outside of the cloud hierarchy (e.g., from the user's laptop over the internet) additional mechanisms and controls need to be put in place for controlling IAM access permission boundaries, which is beyond the scope of this paper.

Finally, there are a couple of additional principles that are cornerstone to protecting applications in the cloud and that apply across all the architectures that we will touch on in this paper. These are:

- **Defense in depth** - this refers to the principle of applying multiple layers of security and protection, so that there is extra protection even if a given security control fails to perform its function. An example of this can be to apply a network access control, while an application also performs identity checks to access a given application. Often this is achieved by having multiple overlapping controls at different layers targeting the same problem.
- **Least-privilege access** - this refers to the principle of a security administrator restricting access to the least amount required to perform a job while accessing applications and data.

**Zero Trust security architectures** are built on these principles of defense in depth and least-privileged access. Whereas perimeter security conveys an implicit notion of authority and access, Zero Trust architectures require that all access checks be explicit and done by each application. They do not depend on traditional network controls, and instead, the access to particular applications and data is granted based on identities (ideally digital cryptographic identities, but could use other techniques like tokens). They also often evaluate and map the identity to additional contextual human client or workload attributes, like network IP address for location, type of device, OS, and workload sensitivity. Given this, they can provide a high fidelity assessment with no risk of impersonation as well as multidimensional policy evaluations that enable defense in depth. Zero Trust architectures for cloud [75] are described in detail in Section 3.3. While these Zero Trust architectures like BeyondCorp [15,16] are the north star, they are hard to extend and make work with non-cloud existing environments. In these cases, they can be combined with hybrid service-centric architectures described in section 3.2. It is also worth noting that even in a pure Zero Trust environment, in practice, perimeter security does not entirely go away. The perimeters tend to get much larger and less granular, but do not entirely disappear; while Zero Trust techniques are overlaid to provide application-centric granular controls.

**Section 3: Data Plane Security Architectures**

In this section, we will explore how to use the three key concepts defined in the previous section (perimeters, landing points, and combining with Zero Trust) in cloud application architectures. These architectures are based upon our experience working with security-sensitive administrators and application owners, as well as internal deployments on Google Cloud. For each architecture, we will evaluate their pros and cons, and offer some recommendations on ways to combine them. We will consider three architectures:

1. A **lift-and-shift architecture** whose main goal is to faithfully reproduce the on-premises world in the cloud. This architecture will introduce a network-centric perimeter.
2. A cloud-native **hybrid services architecture** that leverages the landing point and cloud security perimeter concepts.
3. A **Zero Trust distributed architecture** that implements distributed security controls and leverages Zero Trust.

In all variations, controls are implemented at multiple layers of each architecture. In order for cloud application owners to achieve a business outcome, they deploy workloads that contain their business logic and associated data, which needs to be generated, processed, and stored. To get started, we will define a few key terms that we use throughout this paper:

- **Application layer** - the code and data that comprise specific business logic and process. The application is typically front-ended by a set of programmable interfaces (APIs) that are offered to external human clients, and other applications to consume. For example, one such API might be a human-consumed website. An application consists of business logic, data, and network services implemented on the infrastructure layer.

- **Infrastructure layer** - the compute, storage and networking resources that the application requires to run. Traditionally, much security is implemented at this layer because it provides a point of leverage for governing heterogeneous applications and is amenable to mandatory controls that cannot be overridden by application teams.

- **Infrastructure-as-a-Service (IaaS)** - the provisioning of software-defined virtualized private infrastructure. Application owners are responsible, and have controls, to secure the virtual infrastructure layer, as well as their own application layer built on top of it (e.g., VPC network).

- **Platform-as-a-Service (PaaS)** - In this type of service, an application owner is not responsible for managing infrastructure resources, and simply deploys their core business logic and its data. The virtual infrastructure is managed and operated by the cloud provider. Examples include Compute services like AppEngine or Cloud Run and Data Analytics services like BigQuery.

- **Software-as-a-Service (SaaS)** - In this type of service, the cloud client using the application is not the owner of the application, but simply consumes the applications that are offered as a service by the provider, via APIs and application UIs. Access to data in a SaaS product must be managed through solutions based on application-specific controls [44] like a Cloud Access Security Broker (CASB).

In this paper we focus on implementing the same basic security posture in each architecture examined. Our posture will feature two different *abstract cloud security perimeters* based on two different classifications of data sensitivity. Each architecture has a different set of security controls (at different layers of the stack) used to implement the same basic security posture. We will focus on securing the data planes using perimeters and landing points, and the two perimeters we focus on in our examples are:

- **Yellow perimeter -** the posture is to keep workloads in the yellow perimeter from accessing any infrastructure, data, or applications outside of its perimeter.

- **Green perimeter -** the posture is to allow the applications in the green perimeter to access both the internet and yellow perimeter applications (and data, if necessary). This means the green perimeter is a client of the yellow perimeter and is also allowed internet connectivity.

**Section 3.1: Lift-and-Shift Architecture**

In a lift-and-shift architecture, the primary goal is to reproduce one's existing security architecture on-premises in the cloud. In this discussion, we assume a more traditional on-premises network deployment. We recognize that not all on-premises workloads use this traditional model and some have adopted more modern approaches, such as the Zero Trust model described in section 3.3. In both cases, the techniques of section 3.4 can be applied to combine these architectures within a single cloud deployment as needed.

In the lift-and-shift approach for a traditional on-premises workload, cloud-native security concepts are used at a minimum, except when cloud-native services (e.g., blob storage) are introduced into a workload. A traditional security architecture starts with attempting to keep attackers off the network and then if attackers breach the network, confine their ability to move laterally as much as possible. This is implemented primarily by defining perimeters using networking controls in the cloud that regulates communication between trusted and untrusted domains, but also by incorporating controls at the identity and access management layer.

**Section 3.1.1: Traditional Perimeters for Lift-and-Shift**

In this section we will cover how the network perimeters are typically defined and implemented in a lift-and-shift scenario, in which the user migrates existing on-premises workloads to cloud without significant re-architecture. In this scenario, the network security perimeters in cloud follow and resemble the ones defined on-premises, and they are typically implemented with separate network segments in cloud, mimicking the on-premises properties.

In the following example in Figure 1, the user configures in the cloud two separate network segments, the green and the yellow, for workloads with different levels of data sensitivity. The cloud security perimeter yellow, contains Payment Card Industry (PCI) Data Security Standards (DSS) [76] compliant workloads with sensitive data, like payment processing workloads, with access to stored cardholder account data; while the cloud security perimeter green contains non-PCI workloads that do not store or process any payment sensitive data.

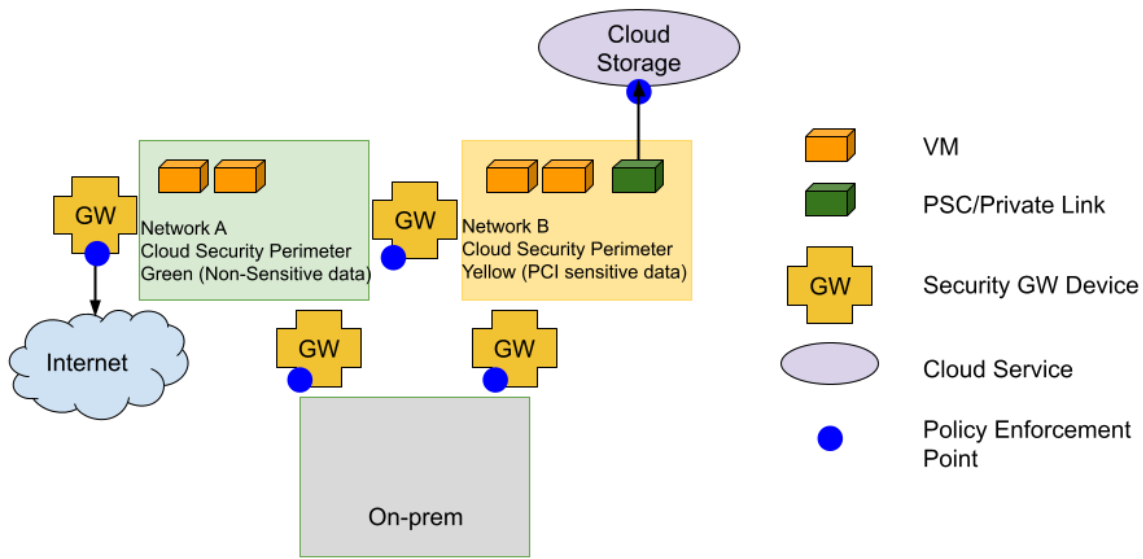

Figure 1: Lift-and-Shift Architecture.

Each of the green and yellow cloud security perimeters in Figure 1 is implemented as a distinct VPC network, but with a shared non-overlapping private address space (typically using RFC1918 IP addresses) that is routable from the on-premises network. The blue dots and the perimeter boxes show the Policy Enforcement Points in this architecture (the security gateways and VPC network boundaries).

Any communication across the perimeters is subject to controls in these policy enforcement points. In addition, this example uses Cloud Storage, for example for storing payment account data in the yellow cloud security perimeter. The Cloud Storage access from the workloads within the yellow perimeter can be provided by Private Service Connect (PSC) [47] and AWS Private Link [48]. Any other workloads outside the perimeter yellow, for example in the green perimeter, cannot access the sensitive data in Cloud Storage, unless it is through the security gateway where controls are enforced.

In this architecture, the network provides reachability between infrastructure entities, for example hosts, machines, servers, gateways, and machines/nodes on top of which applications are running. The infrastructure connectivity networking is Layer 3 (IP) and Layer 4 (protocol and ports). The protection controls typically include:

- **Network filtering and access control based on L3 and L4 attributes (i.e., IP addresses, protocols, and ports).** For example, enforcing IP allow-lists and deny-lists to restrict requests from certain IP addresses or geographical regions, or restrictions to accept traffic only destined to a specific destination port.

- **Volumetric protection, based on connection volume and patterns, to protect against Denial-of-Service (DDoS) attacks from the Internet**. The volumetric DDoS protection can be implemented at Layer 4 network and below.  This helps protect against attacks, such as SYN floods, IP fragment floods, and port exhaustion.  Even so, where possible, it is more effective to provide protection at Layer 7, where it is possible to protect against HTTP request floods or attacks on DNS services, instead of simply bytes. The volumetric protection can be combined with Web Application Firewalls that include application level logic and signatures to identify attacks like SQL injection, PHP injection or cross-site scripting [49]. The DDoS protection and Web Application Firewalls (WAF) are very important but complex subjects, and the details are outside of the scope of this paper.

These controls are used to limit flows, both within each perimeter, as well as between perimeters. Enterprises leverage network firewalls and security appliances of varying sophistication, ranging from traditional 5-tuple based firewall rules to more scalable zone-based firewalls and layer 7 firewall products or technologies like global rate limiting and intelligent denial-of-service protection that can analyze and react to requests. Note that despite this sophistication in terms of describing the environment, in many cases these techniques ultimately map back to flows defined in terms of 5-tuples (i.e., L3 and L4 identities). The higher level of abstraction is used for management, but not necessarily in the data plane.

In some cases, these firewall products do operate at layer 7 in the data plane. For example, they can open up and inspect TLS traffic that crosses perimeters for data exfiltration protection reasons as well as provide hooks for implementing intrusion detection and intrusion prevention systems (IDS & IPS)[1]. These firewall configurations can be extremely complicated in a large enterprise network [7].

The gateway devices in Figure 1 are sophisticated layer 7 firewalls. As an example, they may allow new TCP connections from green perimeter to yellow perimeter, or green perimeter to internet, but not from yellow perimeter to green perimeter or yellow perimeter to internet. They would also look for sensitive data moving between the yellow perimeter back towards the green perimeter based on rules analysis. Though cloud platforms provide some of these capabilities natively, these gateways used for lift-and-shift are often provided by a partner in a cloud environment and are similar to the physical devices that are installed in an on-premises network, so that the security in cloud can be managed with the same solutions and interfaces as in on-premises environments.

---

[1] Note that these devices, if poorly managed, can actually make security worse by introducing vulnerable legacy cryptography, leaking sensitive key material, and other similar threats. Whether inspection makes sense is a nuanced topic and expertise on securely managing keys and keeping these systems up to date is critical to success [62].

At the platform level, network security constructs such as VPC flow logs and packet mirroring should be enabled in order to collect additional visibility into network behavior. These controls backstop the security gateways by providing additional visibility into packet flows within the network segments. Such a lift-and-shift architecture can be deployed either within a single region (and replicated in other regions) or where each network perimeter spans multiple regions.

Finally, note that this architecture often relies on the Shared VPC model [58,74], in order to allow the separation of concerns between network administrators and DevOps application owners. In Shared VPC, one central cloud project[2] hosts the VPC network and it is operated typically by the central security and networking teams, through IAM roles on that central project (IAM is discussed in more detail in Section 3.2). Other cloud projects host application workloads, and are owned by application owners who can leverage and connect their application workloads to this Shared VPC running in a central project. In this way, the application owners are empowered to start workloads on the network and have some level of permissions to manage an individual subnet without having full firewall permissions or permissions to change settings (e.g., to change routes). Depending on which permissions are delegated to development teams, different levels of centralization can be obtained. The architecture described in Figure 1 is still applicable in this case, with the green network A and yellow network B hosted in network-admin administered cloud projects, while the VMs are part of different administrative projects, owned by the DevOps teams.

In many cases, the workloads are migrated from on-premises to IaaS infrastructure (VMs), and once such a migration is in place, they can slowly incorporate concepts and technologies from a Hybrid Services Architecture to help secure them, which is described in Section 3.1.3.

**Section 3.1.2: Identity for Lift-and-Shift**

This section covers the aspects on how to extend the identities across on-premises and cloud in the lift-and-shift architecture, so that the users and workloads on-premises can be authenticated and authorized in their access to cloud.

Although the network is typically one of the primary tools for enforcing security policy and boundaries in the lift-and-shift architecture, it should be supplemented with identity-layer security. This requires the identities to be able to be understood across on-premises and cloud, which is the focus of this section.

Typically, a lift-and-shift architecture seeks to preserve the on-premises identity equivalents (e.g., lift-and-shifting the Active Directory server), and this needs to be integrated with cloud identity. Another area of integration is that the Identity and Access Management (IAM) is modeled differently by different

---

[2] A cloud project is a fundamental grouping construct on GCP. It is similar, though typically more lightweight, than *accounts* or *subscriptions* on AWS and Azure respectively. For our purposes, the important point is that typically projects collect related services and workloads that have a single administrative domain. For example, a single DevOps team may manage all workloads in one project and they all belong to the same environment (e.g., dev, staging, or prod).

Operating Systems and application stacks. These IAM systems are unique to those platforms and do not trivially interoperate with the identity systems in the cloud platforms themselves. For example, POSIX uses users, groups, and privileges to model IAM, whereas Windows uses Active Directory. There are technologies to bridge these differences, for example by federating Active Directory with cloud platform identity infrastructure or by synchronizing platform identity into unix identity, but these technologies are bespoke and in a lift-and-shift world it is typical to treat the world of cloud platform identity and application layer identity separately. That is, the identity infrastructure from on-premises (e.g., Active Directory) is copied into the cloud and not integrated with the Cloud Service Provider's identity infrastructure. Patterns for how to integrate lift-and-shift identity with PaaS services are described in Section 3.2.3.

As its name suggests, IAM can be thought of as being divided into two major areas:

- **Identity** - The *Identity* area concerns modeling entities that are relevant to security goals with a robust set of abstractions, issuing credentials that represent those abstractions, and governing the lifecycle of those entities. The phrase Authentication (AuthN) is often used to capture some of these concepts.

- **Access Management** - The *Access Management* area concerns managing what those identities can access. The phrase Authorization (AuthZ) is also used to cover these ideas.

The identity management area is extremely broad, so for the focus of this paper we are going to narrowly focus on identity issues related to protecting the administration of the workload and the access to its associated code and data. To do this, the first step is to enumerate what parts of the application need identities. In this architecture, identities are required for at least the following key objects:

- **Raw workload entities** (e.g., IaaS resources like VMs or auto-scaled k8s pods, PaaS resources like serverless functions or data views) - these use identities to authenticate to each other and to cloud services.

- **End human or application clients who interact with an application** - many applications require different access for different people or types of client applications.

- **Developers or operators who administer an application** - different access rules allow different people different degrees of access to administer a workload. This is typically achieved through Role Based Access Control (RBAC) [50], which restricts access based on a person's role within an organization.

Identity is usually managed by an Identity Provider, which is a service that is responsible for allocating identities to objects (e.g., components of a workload or a person), managing the lifecycle of these identities, and issuing associated credentials (e.g., tokens, X.509 certificates [63]). In the lift-and-shift architecture, typically an enterprise has procured one or more identity management solutions from a vendor (e.g., Okta, Microsoft Active Directory) and seeks to make this Identity Provider interoperate with their cloud platform and continue to function for workloads moved to the cloud.

Once identities are defined, their access to resources must be managed. Various techniques have been developed over the years such as Role Based Access Control (RBAC) and Attribute Based Access Control (ABAC) [50]. Our paper does not go into these in detail, but we note that in a lift-and-shift architecture, one traditionally preserves the approach to identity that was used on-premises. Often this involves replicating the identity provider used on-premises, such as Microsoft Active Directory, in the cloud either using IaaS or a PaaS managed version and then *federating* this cloud-based identity provider (IdP) with the on-premises IdP to enable applications to continue to function.

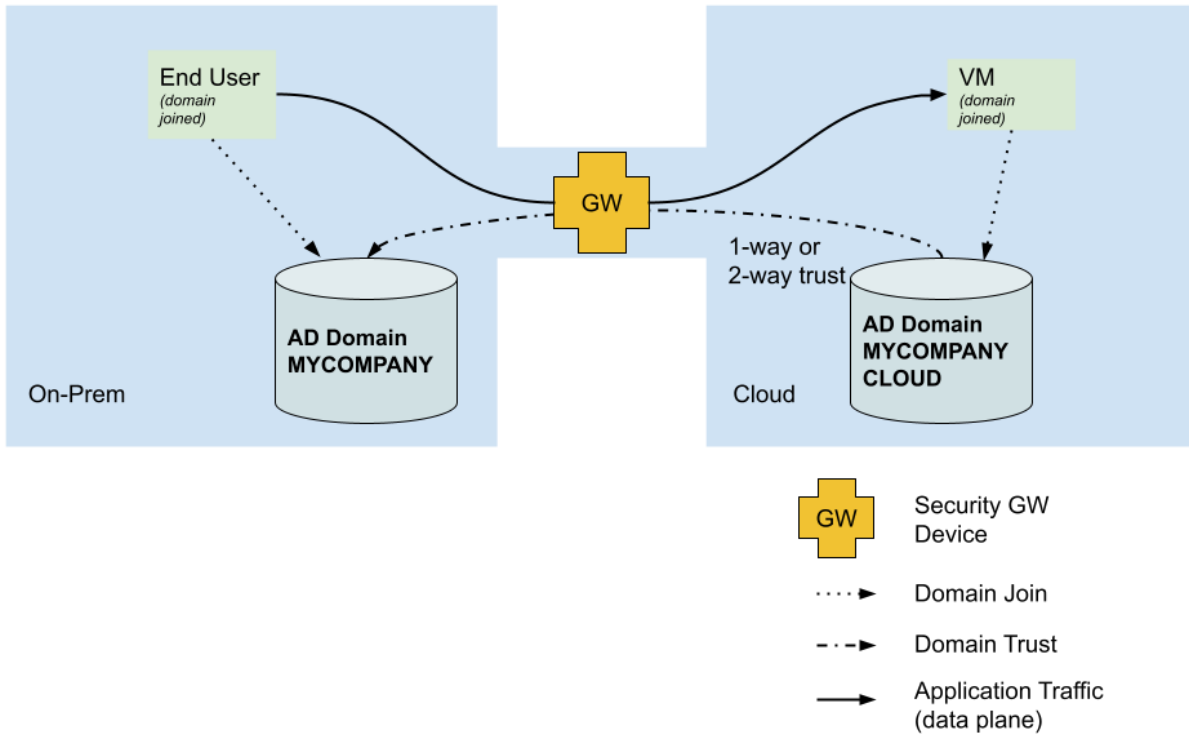

Figure 2: Federated Active Directory Environment.

Figure 2 shows an example topology where the cloud footprint of Active Directory (AD) issues credentials (kerberos tickets) to workload infrastructure (machine identity) and services that are being run. In this type of deployment there are typically two AD domains – one in the cloud and one on-premises. The cloud domain *trusts* (a form of federated identity) the on-premises domain, meaning that it will honor credentials issued by the on-premises domain. That way, people (both developers and end application clients), who have identities issued by the on-premises Active Directory, can leverage this AD trust to authenticate to the applications in the cloud. Cloud applications have identities issued directly by the cloud domain and can authenticate to each other. In some cases, there is also a reverse trust or a two way trust [64] to enable on-premises applications to trust identities in the cloud. This picture is very similar to a traditional on-premises enterprise deployment of identity with multiple domains that trust each other. Direct network connectivity at layer 3 is required between the cloud and on-premises environments here because Active Directory does not officially support NAT configurations [39].

**Section 3.1.3: Challenges Incorporating Cloud Services into Lift-and-Shift**

The introduction and use of Cloud Managed Services, which are cloud-native in nature, complicates the lift-and-shift architectural principles, both for identity as well as for networking as follows:

- **Identity** - As described in previous sections, the approach to separate the legacy identity for applications and the newly introduced cloud infrastructure identity is one that is often used in lift-and-shift architectures. This approach is not possible anymore when using cloud managed services that use native cloud platform identity. This makes it hard for legacy applications and cloud managed services to authenticate and authorize each other.
- **Network** - Similarly the network access paths for Cloud managed services are implemented and controlled by the cloud platform provider. The routing is often done at application level routing, with the cloud provider managing the underlying network infrastructure, which creates challenges for controlling access using traditional on-premises networking approaches.

Let's look further into these challenges.

In terms of identity, in order to create compatibility and authenticate on-premises applications, lift-and-shift applications get identities that have been asserted at the application layer by a lift-and-shift Identity Provider like Active Directory. In addition to this, the cloud provider itself acts as an Identity Provider, since the Cloud platform typically asserts identities for workloads that are used to authenticate to cloud APIs or even to other workloads. These cloud-native identities are typically provided by a *metadata server* that can be used to obtain tokens for the cloud identity of a compute resource and used to authenticate to cloud resources or other applications. For example, they can be used to authenticate cloud native services like blob storage, even when creating a lift-and-shift architecture. However, these cloud-native identities are independent from the identities used by the AD infrastructure. Therefore, a given VM might have multiple different identities depending on the type of service it is authenticating to. This can pose additional architectural complexity challenges when an application needs to understand both identities from the lift-and-shift identity provider and the ones issued by the cloud provider.

To make the two identity providers compatible with each other, identity such as AD can be *federated* to obtain tokens for calling cloud services by using techniques such as GCP's workload identity federation [40] or AWS's stateful token service [41]. This allows principals that are known only to the lift-and-shift identity provider (AD) to obtain access to systems that only understand cloud-native identity from the platform provider. This type of identity federation is extremely powerful as it avoids the need to store key material in software, which is a security anti-pattern[3], while also allowing infrastructure components to reuse their identity in a diverse set of contexts.

Modern cloud services and applications present various challenges for traditional lift-and-shift network security due to (a) potential IPv4 space exhaustion and managing a non-overlapping routable IP space, and (b) due to securing and restricting the access to *dynamically created and non-persistent* backends and their IPs using traditional network perimeter controls. Let's now explore these challenges in specific cloud-native technologies like Kubernetes or serverless applications.

Kubernetes has the challenge of IPv4 exhaustion in lift-and-shift architectures as many containers/pods are packed into VMs. As IPv6 is adopted this would not be an issue [84], but in IPv4 environments, large Kubernetes cluster network configurations require the reuse of IPv4 address space with non-routable networks. In addition, Kubernetes is often focused on stateless applications, where IP addresses for containers are dynamically created and removed as containers are dynamically managed for the running service. This means as the containers come and go, a traditional lift-and-shift network security perimeter would need to be updated.

---

[3] A growing volume of sensitive data - or secrets – such as API keys, private keys, certificates, username and passwords end up publicly exposed on GitHub, with 20% growth year-over-year [61].

Serverless technologies like Cloud Run and Cloud Functions produce ephemeral compute entities that continuously recycle internal IP:ports, which are internal to the cloud provider's infrastructure. While these technologies do not have the challenge of the user IP space allocation, providing secure access in lift-and-shift architectures requires additional integration. These functions are addressed at the application layer with a Fully Qualified Domain Name (FQDN) [66]. Then to integrate them into a lift-and-shift architecture, for Google Cloud Functions, **VPC connectors** [51] bridge the traditional network controls with the application layer Cloud Functions. VPC connectors are able to provide private network connectivity and access controls to the cloud function in these lift-and-shift architectures. These VPC connectors make it easy for the cloud function to be part of a VPC, abstracting the complexity of the non-persistent compute resources, as well as providing additional network controls that can prohibit the access from the function to external networks. Other clouds also have mechanisms to make their serverless technologies work in a lift-and-shift world. For example, AWS uses Network Address Translation (NAT) and cross-account virtual network interfaces to allow Lambda functions access to VPC resources [81]. These mechanisms inherit the limitations of network-centric security common in lift-and-shift deployments, creating an impedance mismatch with the flexibility and scalability these services generally offer. For example, implementing different security boundaries and policies for different functions requires understanding and tracking the mapping between their logical identities as functions—which are often ephemeral—and the "physical" identities of the VPC connector or NAT-ed network interface—which may be at a different granularity. This complexity challenges the simplicity and dynamism of the underlying serverless technology.

As an example, the function's specification (not just its execution) may be rapidly evolving because a data engineer is frequently editing it, where they also need to think about FQDN/IP address planning and network security. Imagine that you have some functions that are used in a data pipeline. You create a bunch of different copies of the function to handle different data sets, move data around, call various services, and more. Not only the life of the function is ephemeral in this situation, but also its specification in the control plane--new types of data sets can come relatively often and the data engineer goes in and updates parameters of the function for the new data set. Because the function is so lightweight, it is useful in these kinds of situations versus more complicated tools. For this example, to define security in terms of network policies, you need to think in terms of how those functions assume the IP addresses of the various subnets/VPCs that call the services and secure them. Basically, it requires a lower level network reasoning about the security boundary in terms of concepts that are less flexible than the function (e.g., you need to plan network address space to align with your security boundaries, you need to have enough distinct addresses/subnets to align to the different security boundaries of your different data sets that you might want to create functions to talk to).

In order to address these challenges, bridging abstractions are necessary that connect the more modern cloud centric concepts to the network infrastructure. These abstractions form the foundation of "service networking", which lead to an alternative approach for protecting cloud applications in the landing point architecture described in Section 3.2.

**Section 3.2: Hybrid Services Architecture**

An alternative to the lift-and-shift approach is to move more towards a cloud-native networking and identity model. We will start by introducing Cloud Security Perimeters and how to apply them to secure network boundaries. We will then introduce the concept of a Cloud Landing Point, which is a configuration of the cloud infrastructure that heterogenous types of workloads can be plugged into in a relatively uniform manner. In addition to Cloud Security Perimeters and Cloud Landing Points, a core concept we will use is what we call service networking, which is a form of application-layer networking and cloud native identity. By **Service Networking** we mean a networking abstraction that represents a set of backends serving a common application. The access to the set of backends that serve the application is through a service identifier represented by an IP:port (Layer 4) or a FQDN (layer 7), rather than directly to the service backends.

**Section 3.2.1: Cloud Security Perimeters**

In the lift-and-shift case, the cloud security perimeter is aligned to network boundaries, with each segment separated by gateway devices as in Figure 1. That is, the perimeter is implemented using infrastructure networking. As observed in that architecture's discussion, the introduction of cloud native services (both IaaS and PaaS) poses several challenges to that model. These challenges include:

1. **Multi-tenancy** - the boundary between different tenant customers of a cloud service is frequently a logical one implemented inside the service, rather than codified in network firewalls.

2. **Direct integration** - cloud services often talk to each other directly within the cloud, creating new data flows that need policy control based on the application's network, but also identity and other contextual application attributes.

3. **Higher level data model** - the data model of these services often works in terms of service-specific resources and objects that interact with each other rather than lower level network entities more amenable to traditional perimeter control.

In order to implement a perimeter in such services, new approaches are necessary. One approach is to create virtual network endpoints that represent the logical entities in the cloud native services [8, 9]. For example, creating a virtual network endpoint that represents the service (e.g., Google Cloud Storage) using GCP's Private Service Connect or a particular resource in a service (e.g., a blob bucket in AWS S3 fronted by VPC Endpoints and PrivateLink). This is shown in Figure 1 by the small dark green cube fronting the PaaS service like cloud storage. The idea is that instead of having a public, internet routable API to all resources in that service, a network endpoint is created inside the VPC network that routes to the service, or in some cases a subset of resources within the service. This mapping approach has the benefit of compatibility with more traditional network security models, but it does not address the direct integration between cloud native services. The network routing and paths to access these cloud native services are often controlled by the cloud provider, with the application owner having limited control or visibility of the physical paths that the traffic takes. Hence applying application oriented network policy to such flows is challenging.

To address this, Cloud providers have realized the need for a cloud-centric model. That is, allowing the definition of a logical perimeter around the *native cloud data model* as well as *contextual* access rules that combine network-centric and identity-centric views of the world. By *native cloud data model*, we mean abstractions leveraged in the cloud REST APIs like blob buckets, projects, folders, accounts, or resource groups rather than network-focused concepts like IP addresses, or application layer 7 objects like paths and URLs. In multi-tenant PaaS services, these entities increasingly appear in data plane interactions. For example, Google BigQuery allows queries to be run against Object Storage systems without any low-level networking implications. The security boundary in such cases must be implemented within the native cloud data model.

Cloud providers offer constructs to achieve this like GCP's VPC Service Controls (VPC SC) [6]. VPC SC allows the definition of a logical perimeter around the *native cloud data model* as well as *context-aware* rules that combine network, identity and cloud resource hierarchy (e.g. projects, folders) attributes. In VPC SC, communication is allowed within the perimeter (made up of cloud projects), whether from IaaS resources like VMs to cloud native APIs or between the cloud native APIs themselves. To achieve context-aware rules, VPC SC also has a rich rules language for governing cross-perimeter access for both users and workloads, that can evaluate network attributes like IP address and private virtual networking as well as client user identity and device data. Together, these enable the specification of powerful virtual cloud security perimeters around cloud native infrastructure and data resources.

However, such a data plane firewall on its own is insufficient to fully govern PaaS resources, because it only provides a coarse layer of perimeter security. On GCP, VPC SC is complemented by other controls to complete the formation of the cloud security perimeter abstraction. Most of these controls are defined in terms of the native cloud data model. The resource hierarchy is a critical piece of this data model and is a key tool for implementing a logical perimeter boundary around related workloads. The resource hierarchy lets a resource owner group projects (the main grouping unit for resources in GCP) into Folders and an Organization (of which there is typically one per enterprise organization). These folders can be used to group related workloads (captured in development team owned projects) together and apply similar policies to similar types of workloads. Folders can be nested to build out a hierarchy. Figure 3 shows one common design pattern, which is to align cloud security perimeters to folder boundaries and shows how to group different workloads using the resource hierarchy. In Figure 3 we have both the green and yellow workloads as in Figure 1. In Figure 3, workloads are grouped along two dimensions: prod versus development. This is achieved using a nested folder structure, where we separate out production (prod) and development (dev) folders each with their own security perimeters.

This alignment of the hierarchy enables the use of various security controls to realize the cloud security perimeter boundary. All clouds offer some form of hierarchy. On AWS, Organizations own and help manage Accounts [52]. On Azure, Organizations own Subscriptions, which help manage Resource Groups [53]. Both Azure and AWS also provide various degrees of policy controls that can be managed using their hierarchy to achieve similar outcomes. In this paper, we discuss in detail the controls in GCP that should be used. While the cloud hierarchy is used to define cloud security perimeters, there is a need for additional flexible groups where policies apply. This is because every resource can only be in a single hierarchy and policies like compliance, rollouts, or billing, often need to apply to groups of resources that do not always match the hierarchy. To address these problems, all three major clouds also support "Tags" or "Labels" [54,55] to enable arbitrary grouping of resources without a fixed hierarchy. Tags create slices "across" groups of consistently labeled resources. Tags are more flexible, but also more complicated, than the hierarchy and are often employed in these cases where there are multiple independent and overlapping classifications to apply to resources with different ownership. In some cases, they may also be used to define a cloud security perimeter as well, in lieu of the hierarchy. In other words, the cloud security perimeter could be created as a combination of the resource hierarchy (for example a set of specific folders), and a set of tagged resources or a list of cloud native services defined across that specific set of folders.

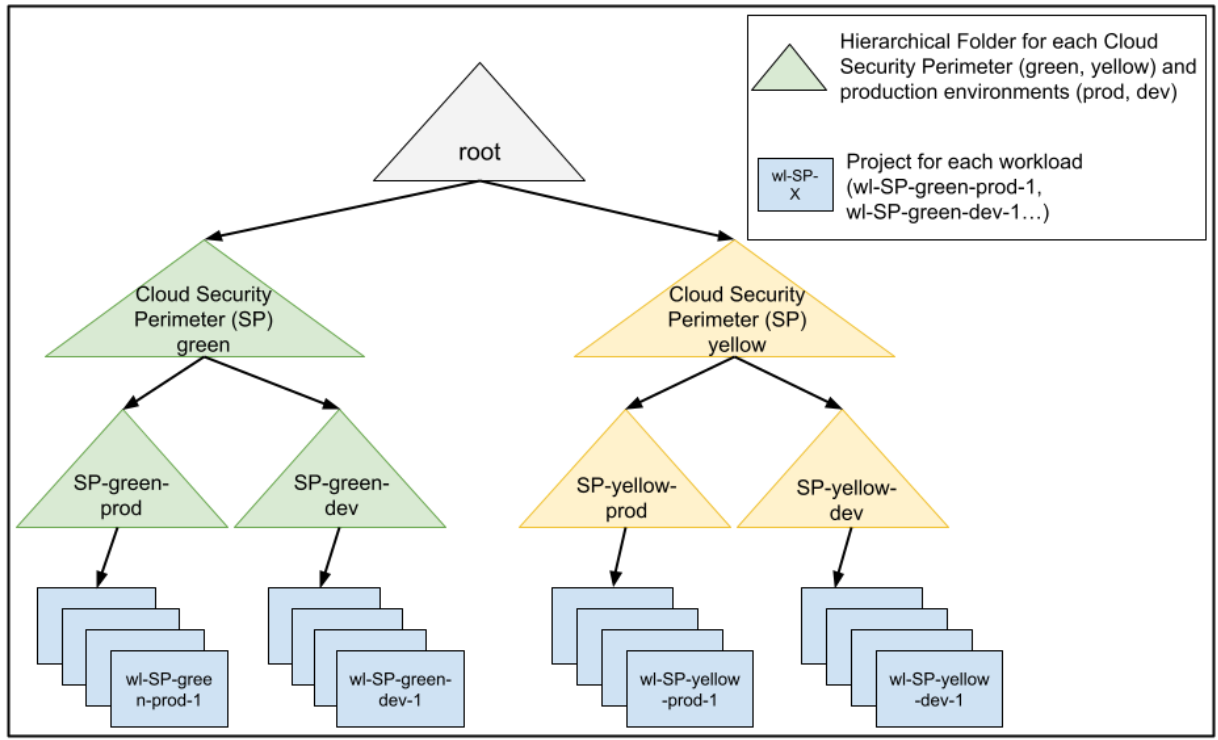

Figure 3: Usage of hierarchy to define perimeter boundaries. Projects are grouped by cloud security perimeter and environment (prod, dev, etc) in the hierarchy.

Best practices for achieving a cloud security perimeter are:

1. **Project and network isolation** - Each workload is deployed in a different project with a non-routable network per workload that will be managed by different development teams, with distinct IAM roles. The details of this are discussed in Section 3.2.2.

2. **VPC SC** - Each cloud security perimeter (green and yellow in our examples) has a different VPC Service Controls perimeter containing the projects for all of the workloads. Ingress and Egress rules are used to manage cross-perimeter requests by specifying the explicit GCP projects, services, and methods that are allowed to communicate across perimeters.

3. **Use of hierarchy** - All projects in a given cloud security perimeter type should be placed in the same folder hierarchy. On AWS, tags can be placed on accounts to achieve similar outcomes. Azure also has a version of tags to achieve this.

4. **Dependency management** - Services that depend on other services can create reliability and security issues (e.g., by creating vectors for lateral movement). While almost all useful services have dependencies on other services, thinking critically about these dependencies (e.g., are they in the same cloud security perimeter, are they necessary for application function) and managing them is important for building secure applications.

5. **Hierarchical firewalls** - Hierarchical firewalls should be used to enforce a uniform firewall policy for each workload in a particular perimeter. For example, one perimeter may not permit internet egress at all. This can be enforced by placing a hierarchical deny firewall rule [67] on the folder to block all internet egress from projects in that folder, and thus in that perimeter.

6. **Tags** - IAM, together with Tags, can be used to manage fine grained ACL control on access to the individual resources from the workloads on all major clouds. The tags allow flexible classifications that can span across the resource hierarchies, for example, across multiple projects in a folder, or multiple folders in an organization. The tag assignment is often delegated to development teams to apply data classifications to projects that govern IAM. For example, tags might be used to apply data classifications to projects that govern IAM permissions (e.g., allow one project to access storage buckets with tag "pii:false" without granting access to storage buckets with tag "pii:true"). Development teams or data operations employees would be able to manage the tag assignment (e.g., tag as pii:true) separately from the access grant, enabling separation of duties and least privilege.

7. **Organization policy** - Organization policy [68] is used to implement governance and manage the type of resources that can be provisioned into the cloud by its control plane. Although these are control plane policies, they also regulate the configuration and constraints of what is configured in the data plane. These are per-perimeter guardrails that define constraints of the types of resources allowed (e.g., no public IPs, no external Load Balancers (LBs)). These policies are aligned with the folder-based perimeter boundaries but exceptions can be managed using tags (e.g., to allow an external LB in a project within a perimeter that would normally not be allowed). The topic of control plane policies, like Organization policies, is a large topic on its own and is not the main focus of this paper.

The combination of these controls, some in the data plane and some in the control plane, creates a cloud-focused perimeter around data and infrastructure resources in a way that moves beyond infrastructure-networking centric constructs. In AWS and Azure, similar architectures are possible [10,11]. The controls described above are resource-centric, that is, they protect the access to cloud resources and the sharing of its data. Cloud Providers also provide additional Identity-centric based controls that can be applied to provide protection against principal identities in an organization sharing data with other organizations from non-cloud locations. Examples of these controls are IAM Policies on identities (as opposed to on resources), Access Permission Boundaries, and Service Control Policies [69] or Active Directory tenants [70]. Further details on these identity-centric controls are outside of the scope of this paper.

The cloud security perimeter has the ability to include and evaluate attributes that belong to the application and API layers. This means that the perimeter boundary policy can be defined *in terms of the application and API attributes*, allowing for an approach to security in line with Zero Trust principles [2]. For example, the perimeter can evaluate both the identity or device accessing a cloud service, and the cloud resource hierarchy to which the cloud service belongs to. This enables the cloud security perimeter controls to enforce a protection boundary around cloud resources even in the absence of a traditional network manifestation for those resources. This approach is powerful because it enables reasoning about security in the context of cloud native abstractions and concepts, which is typically simpler and more intuitive. Within the perimeter, workloads can adopt different security models, with some leveraging more traditional network based techniques and others leveraging more modern Zero Trust and identity based technologies. We examine this in Sections 3.3 and 3.4.

**Section 3.2.2: Landing points and their network design**

Cloud security perimeters can be complemented by adding Cloud Landing Points (CLPs), which adds a "hub and spoke" network architecture. The cloud landing point network serves three key purposes:

1. **Creates a beachhead in the cloud environment** that is conceptually an extension of the on-premises infrastructure where workloads can be exposed. This serves as a bridge between the modernized cloud and legacy on-premises environments.
2. **Decouples infrastructure networking concerns** like peering locations of interconnects from cloud native application and service architecture concerns. Concretely, thousands of workloads can be developed largely independently of each other, and only deal with publishing themselves into the landing point to be accessible by other services, versus coexisting on shared network infrastructure where developers must coordinate with the central security team in order to deploy security compliant applications. Similarly, the cloud infrastructure team can manage the security of the landing point network separately from the security of each team's workloads, which are hosted in separate networks that can be delegated to development teams if desired. This approach is different to lift-and-shift in that the centralized security management is restricted to the landing point network (instead of all the networks in the organization), where the sharing of the applications can be regulated, providing more autonomy to the development teams to independently manage their individual workload networks.
3. **Enables compute heterogeneity**. Multiple different types of workload with different compute technologies (IaaS, PaaS, even SaaS) can coexist and be exposed to on-premises using a relatively uniform model of *service networking*, by publishing and making them accessible through a Private Service Connect or Private Link, while each workload's internal implementation details are abstracted behind a non-routable network[4].

The critical distinction from the lift-and-shift case is that each workload explicitly defines a set of *public interfaces*, realized as services, which are exposed to the landing point network explicitly (this is part of service networking as defined at the start of Section 3.2). *Everything else is hidden as an internal infrastructure implementation detail* that is not directly visible to the landing point network. In the landing point architecture, each workload is deployed as either a set of PaaS resources or IaaS resources on a separate non-routable network within one or more cloud security perimeters. Each non-routable CLP network provides a baseline security network connecting to these workloads as well as addressing IP address exhaustion concerns from some compute technologies (e.g., Kubernetes). The logical non-routable network around each workload reinforces the security boundary around each type of workload. Then the CLP cloud network and workload non-routable network are linked together for access through Private Service Connect (PSC)/Private Link. Then cloud security perimeters are added on top of this for additional defense-in-depth.

In Figure 4, the Private Service Connect (PSC)/Private Link allows the communication initiated from or through the landing zone to the green and yellow non-routable networks. The reason for using PSC through the Cloud Landing Point is that the CLP serves as the network hub that communicates and enables access to services from on-premises. In addition, there are cases in which a workload in the non-routable networks (e.g., yellow or green) may need to start the communication to on-premises through a workload running in the CLP scope.

---

[4] Other models, such as the Zero Trust distributed services model, can support multiple compute types as well but with different abstractions. However, the hub and spoke model provides a clear abstraction for insulating modern compute technologies from more legacy environments (by abstracting them behind a virtual IP address).

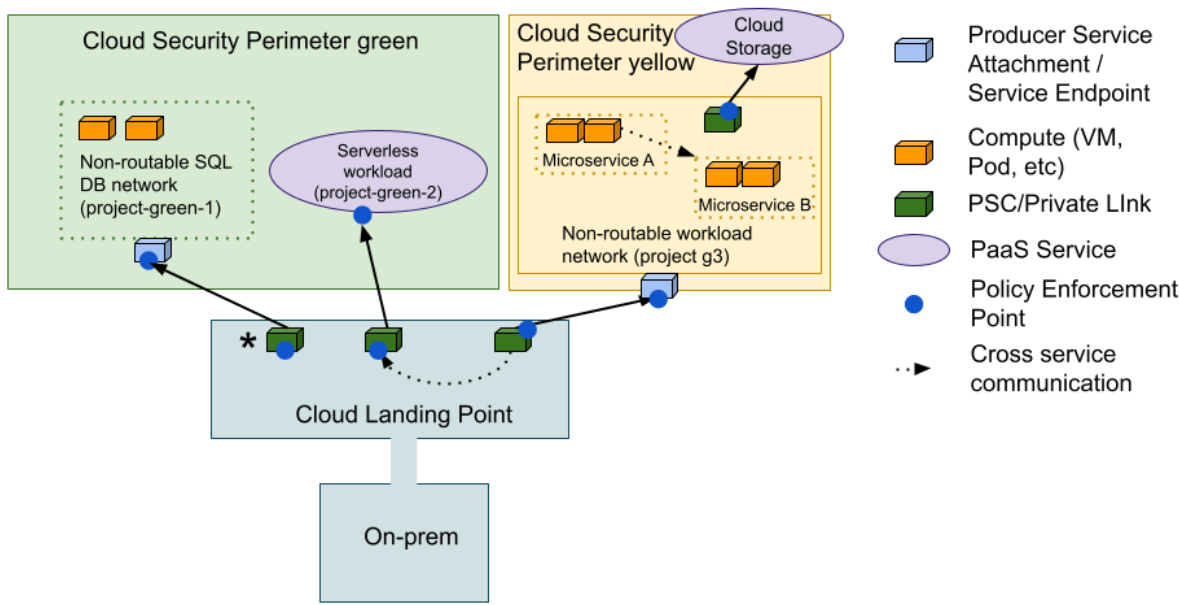

Figure 4: Cloud Landing Point Network Architecture.

Figure 4 shows an example of our two perimeter environments in a landing point configuration. It focuses on the deployment of the network and the core data plane for applications. In the CLP architecture, each workload exports a small, well-defined set of services to the shared CLP network using application networking technologies like GCP Private Service Connect (PSC) or AWS and Azure Private Link as shown in Figure 4. These technologies provide cloud applications with a service network endpoint (either a L7 FQDN or a L3 virtual IP) in the CLP network VPC, while the application backends are hosted in different networks for the yellow and green perimeters. The application service is only reachable through the service endpoint, while the application backends are secured and isolated in a different perimeter. The reachability and access control to the application is achieved through service networking endpoints published in the landing points. For example, an ERP application workload hosted on VMs could expose a single service for talking to the application's HTTP interface. Alternatively, a data processing workload may expose one service which is a set of APIs for submitting new work to be processed and a second service for administrators to configure the workload using an admin web application.

These application networking technologies (PSC, Private Link) provide reachability from human or application clients to Layer-4 and Layer-7 services endpoints. Service networking can operate at Layer 4, where a service is routed to a set of L4 serving network endpoints (IP addresses + TCP/UDP ports). It can also operate at Layer 7, where a service is routed to a set of serving endpoints based on L7 application layer routing, typically including URL paths as part of layer 7 application communication protocols like HTTP, HTTP2, QUIC or gRPC.

This use of application networking (PSC, Private Link) introduces a new control point within the cloud landing point to access the cloud service. The application networking controls overlays on top of connectivity networking. There must be connectivity in order to access the network service endpoint published, but once reached, the service networking endpoint *inside of the workload's cloud security perimeter* provides a control point where there can be additional evaluation based on application layer attributes, like identity or device, in order to gain or restrict accessibility to specific applications. PSC provides two points of control here. First, it provides a published service endpoint for routing and access control to the service in the network of the consumer (in Figure 4, this is represented by the dark green block configured in the CLP). Second, PSC also has a service abstraction in the producer space called the Producer Service Attachment [82,83] (PSA), which is typically implemented by a cloud provider managed gateway. The PSA serves as the target for the PSC/Private Link to attach to the serving application (in Figure 4, this is represented by the light blue block). The PSA is configured by the application owner, who is allowed to configure additional access controls to the backed application to override if necessary the access granted at the PSC/Private Link endpoint in the consumer PSC/Private Link endpoint. Let's look at an example for the SQL Database workload in the green cloud security perimeter in Figure 4. The SQL database backends are in a non-routable network, and the SQL DB workload is only reachable through the PSC/Private Link in the CLP. In this case, the administrator of the CLP can have a policy enforcement applied to the PSC/Private Link, that can restrict access from certain identities or IP addresses (the PSC shown in Figure 4 at the "*"). Additionally, the SQL Database owner and administrator can also apply policies (represented by a blue dot at the Producer Service Attachment for the DB, that can override such access if necessary.

In addition, the PSC/Private Link networking controls provide a consistent method to apply security *irrespective of the compute technology that powers the workload*. For example, WAF, Layer 7 path filtering, and even IDS or IPS can be implemented at the policy enforcement points between the landing point and workload networks. Similarly, identity can be validated and authorized at this Layer 7 interface between the landing point network and the workload, providing a point to enforce Zero Trust security concepts.

Administratively, the landing point network is typically managed by the central cloud team or security team in an organization and is designed to have private IP space allocated that does not need to change regularly. The networking connectivity and cloud security perimeter remains stable, while service endpoints can be easily created to expose new cloud applications into the shared enterprise network, with their associated overlaid control points. The application backend workloads are hosted outside of the shared network and perimeter, and can change significantly behind these scenes without requiring any updates to the security configuration of this network. The IAM roles that manage the project backing this network should be kept closely held and a system designed to govern change requests against this environment. Certain roles (e.g., the creation of new endpoints for services) can be delegated to automation or a less-trusted ops team to accelerate the creation and provisioning of new workloads while honoring the underlying network security posture.

Today, direct communication across the applications in the spokes of the landing point architecture can be challenging because it requires to set up explicit routing and access permissions, for every given direction of the communication across each pair of services. It is possible for such communication to happen, for example, by creating an explicit mesh of Private Link/PSC endpoints between different spokes of the architecture and by using Layer 3 VPC peering as needed. It is worth noting that the complexity in the communication across the spokes can be increased by the use of multi-cloud, given that the user must use different primitives provided by each cloud provider for routing and service networking, as well as, different cloud provided identities, in order to interconnect workloads across each other. There is room for improved platform support across all clouds here, in order to automate, orchestrate and provide easier at-scale management of service networking communication in these scenarios. Given these limitations, CLP is the easiest to use when connecting from on-premises to cloud, where the communication across services is made possible through the CLP as a hub. The next straightforward use is within a single cloud when there is a set of common shared services published and made accessible to applications configured in spoke security perimeters, while the applications in the spokes do not communicate directly with each other.

In addition, for DevOps purposes, direct access to the non-routable workload networks may still be necessary (e.g., for support purposes). Such access can be proxied through a jump box or a product like Identity Aware Proxy on GCP [57], which offers a virtual jump-box-like capability for authenticated and audited DevOps access.

The cloud landing point architecture enables compute heterogeneity. We will now explore additional types of workloads to demonstrate this flexibility. We will explore Google Cloud deployments of VM-based, Kubernetes-based, and PaaS based workloads and how the concepts introduced for cloud security perimeters and cloud landing points apply.

**Section 3.2.3: Compute workload example (VM and Kubernetes)**

Figure 5 describes a CLP architecture with a Landing Point network, which serves as a hub to connect applications from/to on-premises, through network interconnect or VPN. This Landing Point network has published services for workloads in the green perimeter, and serves as a beachhead to provide secure access to those workloads. Although there is a single green perimeter depicted in Figure 5, this is a scalable architecture that allows for multiple parameters to publish services in the Landing Point network.  This allows a workload to have CLPs that act as a network for publishing services to be used across an application or set of applications in the cloud.

We will now go through an automotive business application example, which has two workloads - an Ordering workload and a Shipping workload.   The Vehicle Parts Ordering workload is based on VMs and exposes a service in the green cloud security perimeter. There are two services that communicate with each other as part of this workload - one service is the *Transaction service* and the other service is the *Inventory service*. The *Ordering workload* requires a dependency on object storage, for backup of the data transaction, and has a dependency on managed Active Directory for authentication. The ordering workload has limited egress access to the internet through NAT, in order to update the part numbers as part of the factory catalog. By the policy of the green cloud security perimeter, NAT gateways can be created to allow internet egress. There is a second workload that is the vehicle part shipping portal shown in Figure 5. Many workloads can sit within the same cloud security perimeter as shown in the example, where the vehicle part ordering workload and the shipping portal workloads are within the same green cloud security perimeter. This cloud security perimeter can be governed through hierarchical firewall rules to impose mandatory constraints on developers. In addition to the hierarchical firewalls imposed by the structure of the cloud security perimeters, additional firewall rules can be configured by the workload development team as necessary.

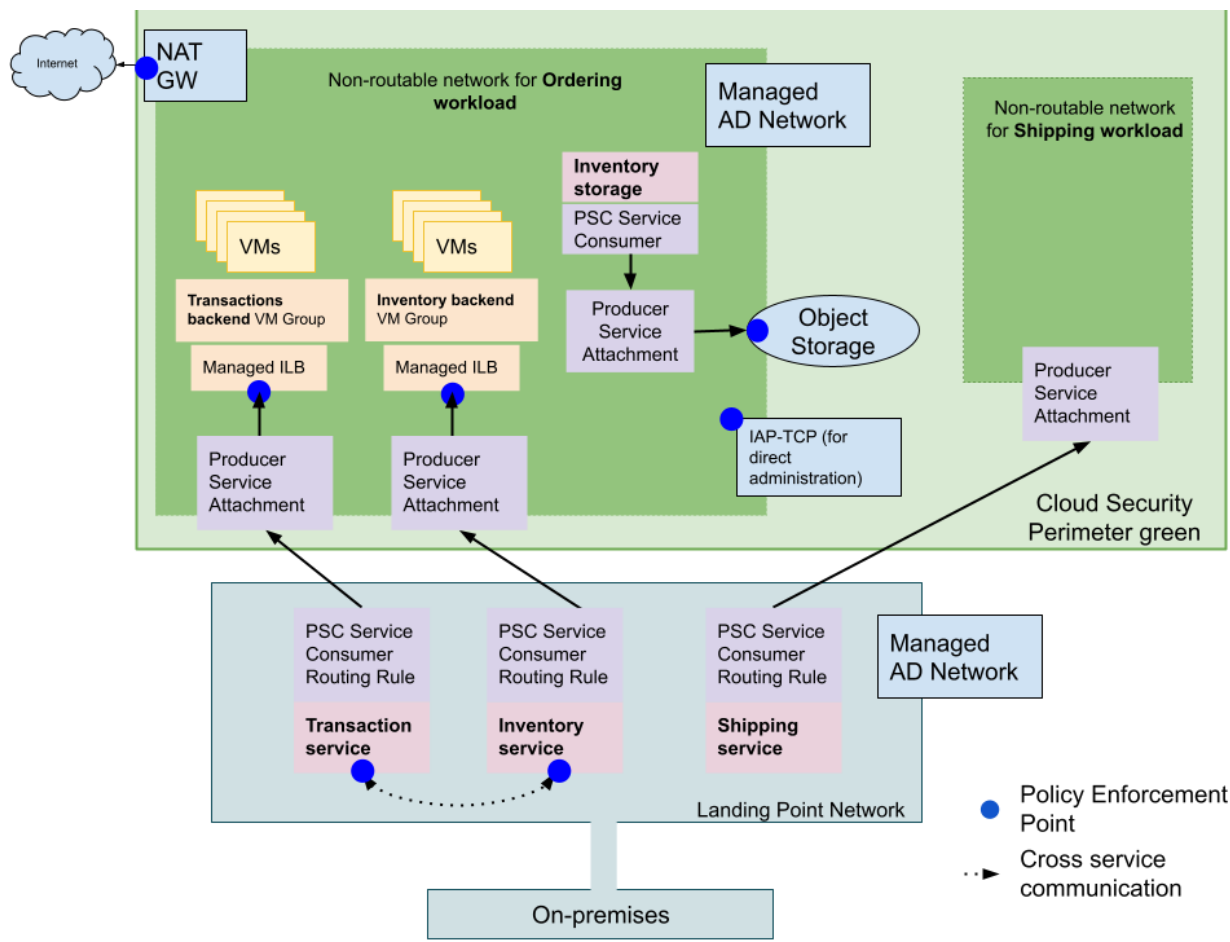

Figure 5: VM Based Service Producer.

There are services that require a flat network to work, even though we have everything set up nicely in this architecture with non-routable networks and a cloud landing point network. Active Directory is an example of a service that requires flat network connectivity with the applications, and therefore network peering is required. If AD is used, direct L3 network connectivity needs to be created between the managed AD service and the network where the applications and workloads are hosted as discussed in Section 3.1. Different cloud providers set up this connectivity differently. For example, on AWS, requestor managed network interfaces are used [12]. On GCP, the AD network is peered with the application network [57]. This peering includes exchanging routes between the networks, thus allowing for packets to flow between both networks as if they were a single network. Given this, it is best practice to host only the strictly necessary AD servers in the AD managed network that is peered with the application security network perimeters. Additionally, network ACLs can be configured in this scenario, denying all traffic communication to the AD peered network except the authorized traffic from/to AD servers.

For most cloud-native applications, the client initiates the connections to a service identifier (IP:port for L7 or FQDN), and it is beneficial to use *service networking* (PSC, Private Link) to access applications across network perimeters. This allows clients in a network access to selected applications in other networks, without having to merge both connectivity domains. Let's consider a cloud native service such as object storage (AWS S3 or Google Cloud Storage). For Vehicle Part Ordering workloads to access Inventory Object Storage, a PSC endpoint with an associated Layer 7 FQDN service identifier is created to model the object storage service as an entity in the green network. Policies can be used to control what data set resources this endpoint may access. These policies vary from cloud to cloud. On AWS, for example, they are called VPC Endpoint policies. On GCS, VPC Service Controls provide the equivalent control.

With respect to identity management, each VM can either use an identity from Managed AD or use its native cloud identity (such as a GCP service account) to authenticate to other workloads and cloud managed services. As discussed in Section 3.1, federation can be used to convert from one type of token (e.g., AD) to another (e.g., GCP-native token for storage). In addition, each workload can offer a different approach to identity under the covers. For example, one might use AD while another uses X.509 certificates and Public Key Infrastructure (PKI) [63]. In addition to the authentication between the end application client and the underlying application, there can be additional identity validation, as well as conversion between different types of identity, performed typically by a proxy between the landing point network and the workload.

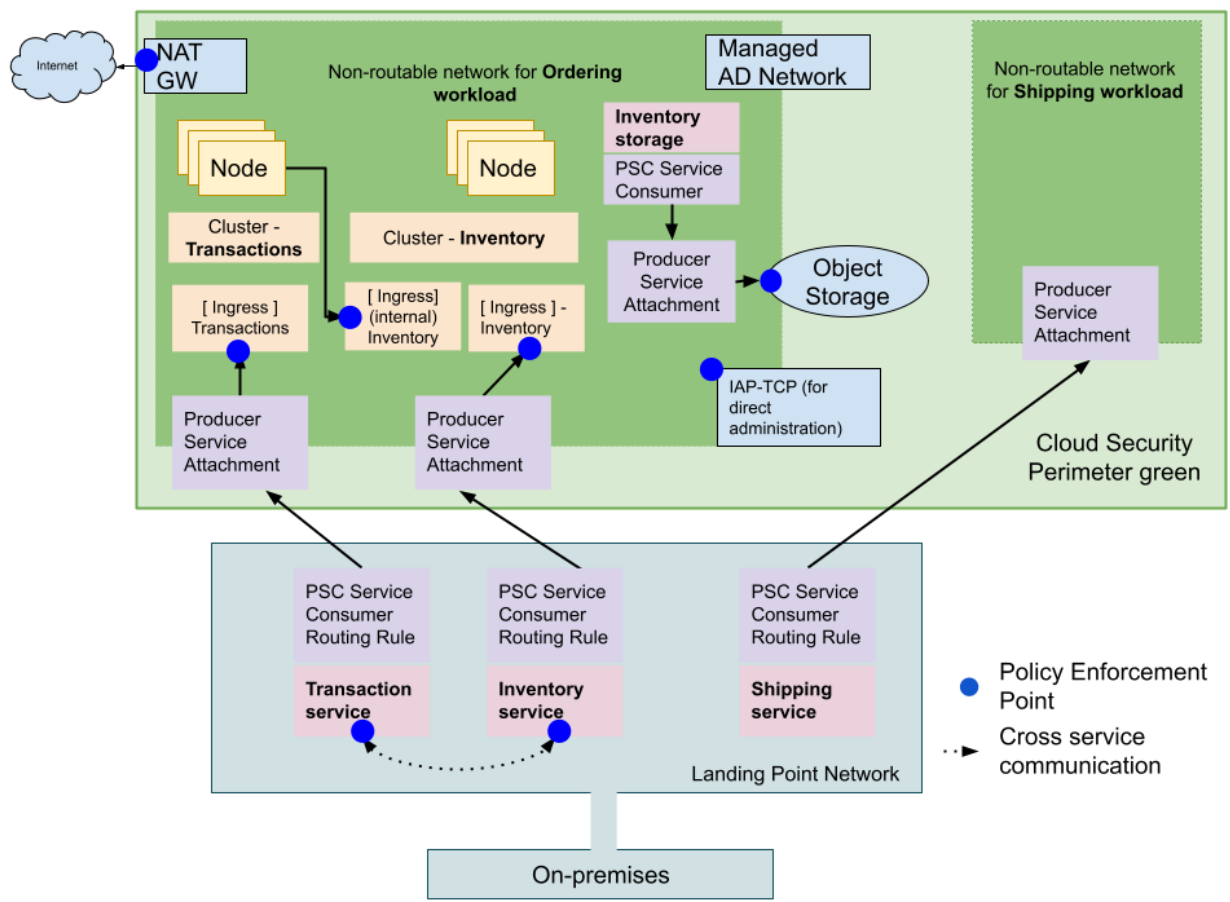

Figure 6: Kubernetes-style Service Producer.

We will now look at the same example but using Kubernetes instead of VMs. Kubernetes allows the packaging of multiple application containers in a single node (VM or server) [77], which creates multiple network addressable pods (one or several containers) per VM [78]. The cluster nodes and the pods are scheduled dynamically, to meet the cluster and workloads declared state, for example, scaling up and down node pools or creating and deleting pods triggered by the creation and deletion of workloads. The Kubernetes services [79] and Kubernetes ingress [72] are stable network abstractions that are used by clients to resolve and route to backends, in this case a set of pods that are serving applications. **Kubernetes ingress** (represented as "[ingress]" Figure 6) provides an L7 FQDN to access the application serving backends, and eliminates the need for direct routability to the backend pods.

Continuing with the same Vehicle Parts Ordering and Shipping example, Figure 6 shows two GKE clusters (Google Cloud's managed Kubernetes offering) that represent the Ordering workload in a distinct non-routable private RFC1918 network [65], one of the clusters exposes a Transaction service, and the other cluster exposes two Inventory Services, one internal and one published to the landing point network. There is also an additional Shipping workload deployed in a separate non-routable private RFC1918 network and that exposes a single Shipping service to the CLP.

Both Kubernetes and VMs can be used to implement the same basic set of services, but some details of the security, networking, and management of the workload itself differ. For example, firewalls work differently in Kubernetes. Kubernetes uses network policy [71] abstractions that are orchestrated and programmed as firewalls in the platform, rather than directly programming the cloud's VM firewall infrastructure or manually crafting IP table rules within the VM userspace. [13] contains a deep dive on Kubernetes container security. It covers critical topics like network policy, RBAC, binary authorization, sandboxing and isolation, automatic node updates, and other best practices. These best practices for Kubernetes security, layer into the landing point architecture outlined previously in an additive way.

In Figure 6, in contrast to the VM case of Figure 5, we create Kubernetes Ingress for accessing the Transactions and the Inventory services, which in turn configures the cloud control plane load balancing. Kubernetes Ingress provides FQDNs in the green perimeter that are locally accessible within the non-routable network, and they can also be published into the landing point network. In Figure 6, the Inventory workload has two ingress FQDNs: one Ingress that is internal, in the sense that it is only accessible within the non-routable network of the workload, and a second Ingress that is exposed and consumable through the PSC abstractions to the landing point network. Similarly, the Transaction workload has a service that is exposed through PSC to the landing point network. In the data plane, there may in fact be multiple proxies here – one on the producer side in the green perimeter (the "service" created by the Kubernetes orchestration layer) as well as a second proxy on the consumer side in the landing point (which may also simply be a routing rule implemented through NAT). This allows policy enforcement points for both consumer CLP admins and Ordering Transaction workload admins.

With respect to identity, on Kubernetes, there is a built-in notion of service account identity that is assigned to each workload that is used both to authenticate within the Kubernetes cluster (either to the Kubernetes control plane or to other workload entities) as well as to PaaS cloud managed services like cloud object storage. For PaaS use cases, these identities can be federated with the platform identity on all clouds and used to obtain tokens for authenticating to cloud PaaS services. Typically, cloud managed Kubernetes offerings include a built-in federation mechanism along these lines. On GCP, it is called GKE Workload Identity [39].

**Section 3.2.4: Data and Storage workload examples**

Data and Storage workloads using analytics tools like BigQuery, Apache Spark, and Blob Storage are common in the cloud. Such a workload can interact with the landing point in several different ways, such as:

1. **Custom Platform with CLP** - One approach is building a platform that abstracts the cloud infrastructure vendor specific details and exposes network services for interacting with the cloud services through a user custom-built platform. In other words, do *not* directly expose the cloud infrastructure even to clients but rather expose it through a custom platform. This option is described in Figure 7.

2. **Direct Access with CLP** - An alternative is to directly expose the cloud PaaS service endpoints in the landing point network and connect to them that way. This option is described in Figure 8.

3. **Direct Access without CLP** - Another option is to leverage the cloud security perimeter without the landing point network, directly accessing the APIs and relying on other controls to compensate (e.g., RBAC on data row as one of many controls). This is described in Figure 9.

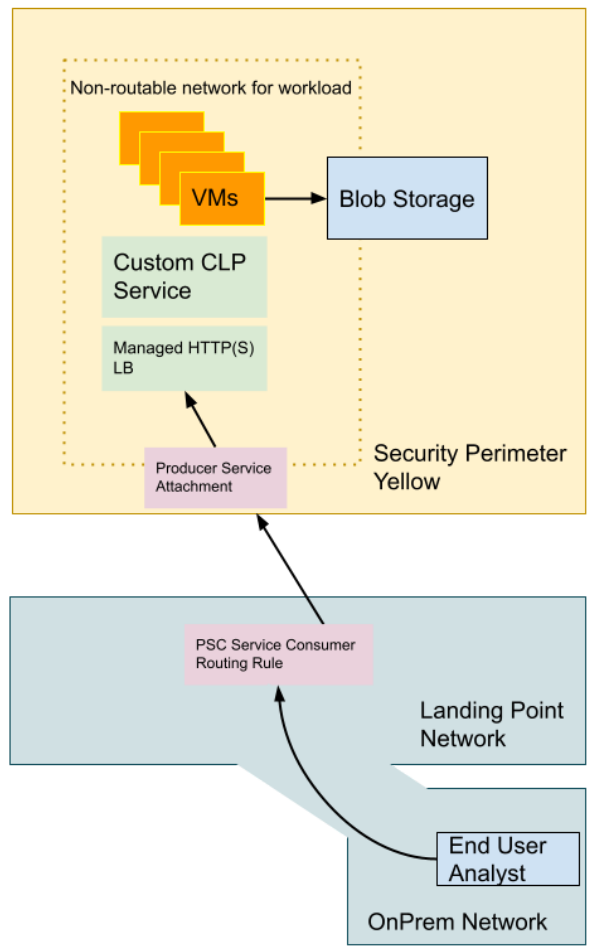

Figure 7: Custom Platform with CLP approach for exposing PaaS services.

Figure 7 describes the Custom Platform with CLP approach. In this example, the application owner that uses the PaaS service within the yellow perimeter, creates their own defined custom service, with a control point that enables authentication and access management. A service endpoint (PSC Service Consumer routing rule) is then published in the Cloud Landing point network, which enables end clients to access PaaS services *through the custom platform*. The access from the end application client is always through this service endpoint in CLP, where additional security access controls are applied. This approach makes sense when one seeks maximum control and when trying to abstract away the underlying service being accessed, but it is an ongoing cost as the underlying cloud service may evolve (e.g., changes of the PaaS APIs) and the custom platform may need to be updated to compensate for these changes.  It is also easiest to apply this approach to services where the API surface area is small and stable (e.g., Cloud Blob Storage).

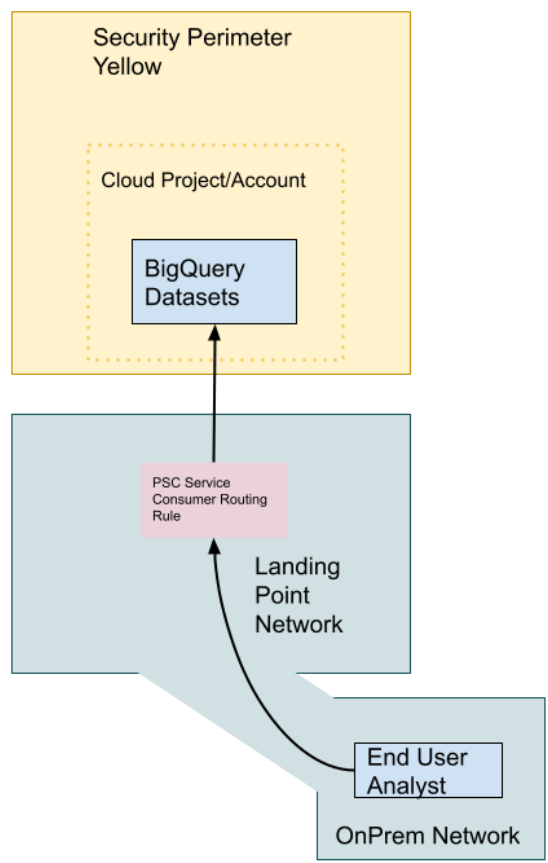

Figure 8: Direct Access with CLP for exposing PaaS data services.

An alternative to the approach of the custom platform is direct access to the service through the Cloud Landing Point. This is described in Figure 8. In a company that is heavily reliant on network based security and private networking, exposing cloud services to end users through this kind of private routing is similar in user experience to an employee running SQL queries against a traditional on-premises database or data warehouse. This approach often eases the transition cost to the cloud and enables a familiar security posture to be adopted, while incrementally layering in cloud native controls.

Figure 8 shows the analyst running queries on their laptop using some business intelligence tool that talks to PaaS APIs such as BigQuery. The connectivity from the tool to the API can be routed through the cloud landing point by exposing the BigQuery service in the landing point network and pointing the tool to that new domain name. This access may be proxied to enforce layer 7 policy, implemented by a reverse proxy [73] that provides API Gateway capabilities, or it may be un-proxied and reliant on cloud platform native policy controls. One downside of this approach is that some software and tooling may have hardcoded assumptions about how to reach BigQuery that require either tricks to work around (such as modifying DNS) or create compatibility challenges that require a different approach. However, many tools enable the flexibility necessary for this approach to work.

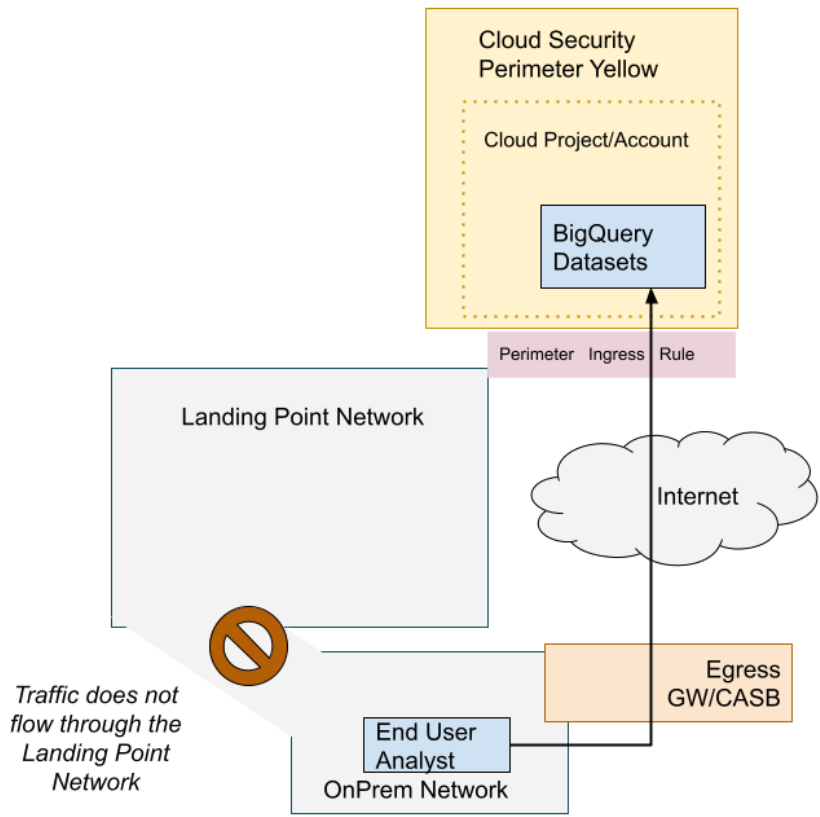

Figure 9: Direct Access without CLP network for exposing PaaS data services.

An alternative approach is to bypass the Cloud Landing Point entirely. We call this Direct Access without CLP network and it is shown in Figure 9. This approach treats the PaaS systems more like SaaS systems, where end application clients directly interact with the platform APIs through the internet (and potentially a Cloud Access Security Broker). This is a viable approach with appropriate controls. Even though the traffic does not flow through the landing point network, there are still cloud-native controls which are part of the platform. These controls serve multiple purposes. For one, they can enforce strong authentication and access policies based on Zero Trust principles even though there is no private network path being used. Additionally, cloud security perimeters can be used to govern how the underlying data can move within the platform (can it be copied between projects/accounts or across cloud organizational boundaries). Together, these mechanisms can govern direct interaction with the platform APIs in either a mandatory or discretionary way to minimize the risk of both unauthorized access and data exfiltration. The protection does not rely on traditional user-managed network or proxy-based controls, but instead all of the policy is managed by the cloud platform.

This direct access approach leverages ideas from Zero Trust by directly exposing the underlying cloud services to the internet and relying on strong authentication and authorization to protect data, together with governance controls to manage other risks like data exfiltration. In the situation described in Figure 9, the end application client is typically a person. In such cases, strong authentication and authorization controls rely on verifying the identity of that person as well as the device they are operating on before allowing direct access. As mentioned earlier, this is similar to how one would secure a SaaS application–less reliant on the network and more reliant on identity to provide security. More details on the required governance controls to comply with this approach are covered in the beginning of section 3.2.

In choosing between the access through a CLP network versus direct access without a CLP network, there are several considerations. The direct access without CLP network is more in line with Zero Trust principles: access directly traverses the internet and security is maintained based on identity not the network. That said, organizations may not have the infrastructure in place to safely execute a Zero Trust approach or the tools to analyze and demonstrate compliance to the security posture, and in that case one of the other options with CLP may need to be used. The CLP network approach is thus often easier for organizations to adopt when moving to the cloud as it is more aligned with traditional network-based approaches. Additionally to the CLP private network protection considerations, there may also be performance and throughput aspects that are made better using private connectivity rather than using the internet for transit.  When using a CLP network, the Zero Trust controls can be overlaid and used in addition to the protection provided by CLP private network perimeter.

**Section 3.2.5: Summary of Hybrid Services Architecture**

The Hybrid Services Architecture explored in this section offers a way of balancing the unique features of a modern cloud deployment with legacy infrastructure. One can think of the Cloud Landing Point as a generalization of the concept of a network gateway to the cloud world in that it acts as a way to manage access to various workloads while enforcing policy. Some gateways, like API Gateways [59], do this at an API layer. Other network security devices (e.g., those discussed in Section 3.1) act as gateways but operate at lower levels of the OSI model. The Cloud Landing Point introduced as part of the Hybrid Services Architecture offers a flexible way of blending these ideas into the cloud.

Similarly, the Hybrid Services Architecture shares some similarities with specific concepts from enterprise architectures such as Enterprise Service Buses [60]. In particular, it incorporates the notion of an architecture based on a set of network services as well as the use of proxies to enforce policy at workload boundaries. What is new in the Hybrid Services Architecture is the flexible abstractions for incorporating a mix of multi-tenant PaaS services provided by a cloud platform with both legacy and modern compute workloads and the mixture of higher level platform and identity based security controls with more traditional network centric mechanisms. This represents an evolution of the architecture for application workloads into the cloud era.

**Section 3.3: Zero Trust Distributed Architecture**

As a final architecture, we will consider an approach that uses distributed enforcement of policies in distributed proxies rather than centralized gateway proxies introduced in Section 3.2. This architecture is used to secure groups of microservices *within a perimeter* that is defined using the hybrid services architecture. From a defense in depth perspective, it often makes sense to embed this architecture inside of the hybrid services architecture of section 3.2, or even the lift-and-shift architecture of section 3.1. We will discuss this combination in Section 3.4, but first we need to consider how to deploy this architecture itself. In the Google production environment [5], there is essentially a single large perimeter over planet-scale services deployments [3,4]. Within this uber-perimeter, Zero Trust principles are used to secure each microservice and enforce policy at the application layer around what identities can talk to each other and from what locations. These identities are authenticated using Application Layer Transport Security (ALTS) [80], which is an authentication protocol very similar to mutual TLS [41]. In this sense, there are "virtual perimeters" enforced at the application layer for accessing microservices.

The distributed (micro)-services deployments are sometimes referred to as part of a service mesh architecture. Such an architecture facilitates service-to-service communication and associated capabilities like traffic management, identity, security policies, encryption, and observability. It is also worth noting that many of the cloud microservices communications extend beyond the cloud to mobile or IoT devices, which in practice must participate in the data plane security. This is typically achieved by extending Zero Trust techniques like device identity or certificates authentication/authorization to the end devices. Additional details about service mesh architectures, and extending data plane security to mobile and IoT devices are out of the scope of this paper.

The distributed microservices approach can be scaled to massive fleets of microservices. The key innovation of this approach is the recognition that within a larger perimeter boundary, it is possible to uplevel security from the network to the application level. This enables the ideas of Zero Trust security to be implemented efficiently. For example, each request between microservices can be authenticated using mutual TLS and authorized using RBAC rules.

More specifically, this model is composed of multiple elements:

- **Application layer policies with contextual attributes** - Typically every service will have an identity. The ability to access the service and the data of a service is expressed in terms of identity access controls policies, but can also include additional attributes that are relevant in the security model, for example, the OS security patching level or the binary integrity of the client service that is performing the access.

- **Distributed enforcement perimeter** - A Distributed Enforcement Perimeter refers to the concept of the Zero Trust policy applied to the granularity of the microservice, which creates a logical security perimeter around the group of compute backends that are part of such a microservice. Each service-to-service communication will have an associated enforcement point. This is typically implemented with a side-car proxy, a middle proxy or a client-side implementation like gRPC. The controls associated typically include authentication, authorization, and encryption, but there are upcoming techniques like WebAssembly [44] that allow the efficient insertion of more sophisticated policies like Data Loss Prevention, that are able to monitor, detect and block the leak of unauthorized data, for example, Personal Identifiable Information (PII) sensitive data, in the enforcement proxies.

- **Logical perimeters for a group of services** - The Zero Trust model offers great granularity to define per-service controls, but this has complexity trade-offs when managing hundreds or thousands of services. In order to simplify the management, the services are often classified in groups (e.g., based on business purposes, data sensitivity, that can be implemented through tags or identity groups) to which consistent protection policies apply. These groups can also be thought of as "logical perimeters", where access across them can be controlled or blocked. By applying the policy to the groups, the number of services can scale with sublinear management complexity.

In this Zero Trust distributed architecture, the application DevOps teams are empowered to apply and manage the controls of their owned applications, but security administrators can still apply mandatory controls by leveraging administrator owned groups and applying mandatory controls to them. This technique for example would allow an administrator to apply controls that verify binary integrity to an admin-managed group of service identities.

Another consideration in this Zero Trust distributed model is that, in order to reason about and audit the actual security properties, the enforcement versus the intended security posture needs to be evaluated across multiple, often hundreds or thousands of enforcement points. This requires a highly scalable cloud architecture for data collection and data analytics, that can collect in real time the controls enforced, as well as can present the right aggregated view that represents the mapping to logical application groups for auditing.

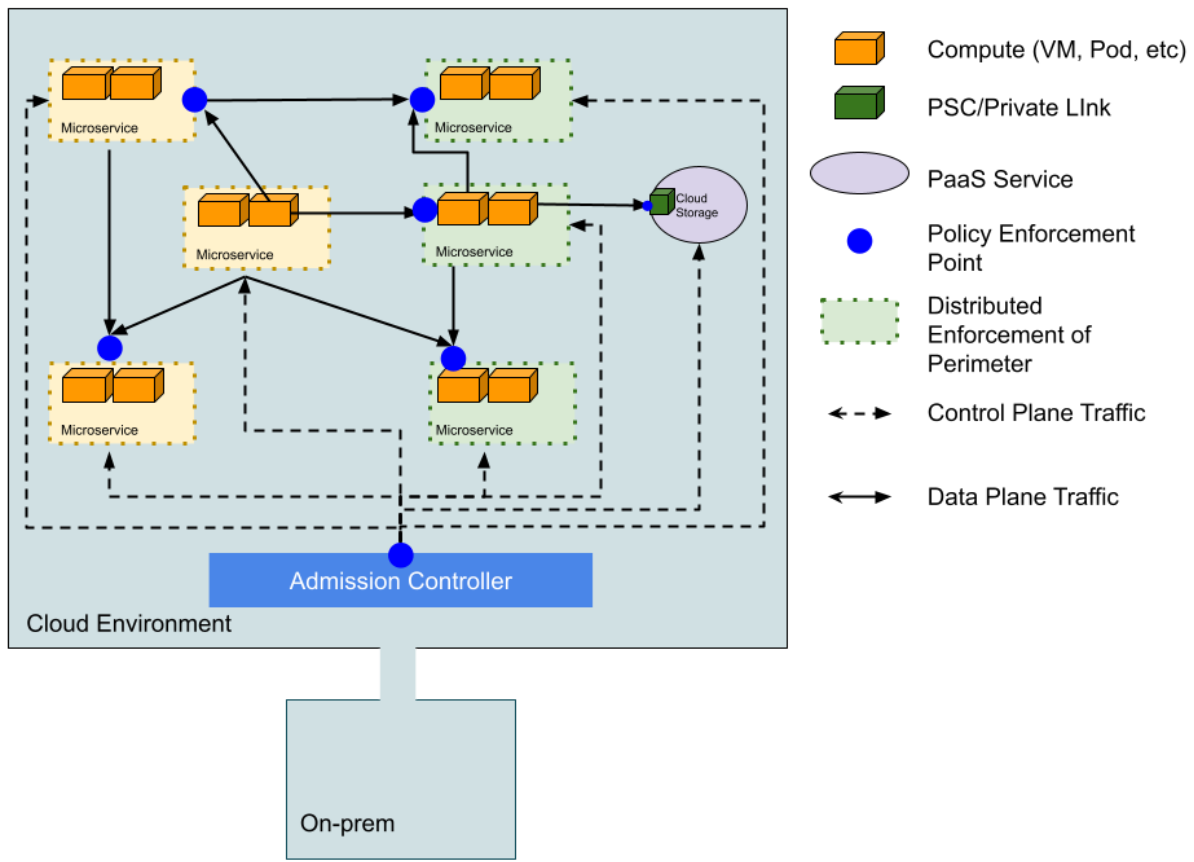

Figure 10 : Zero Trust Distributed Service Architecture.

As described in Figure 10, in this Zero Trust architecture, the "ecosystem" of microservices that can talk to each other is much larger and not constrained by internal network gateway devices [42], DMZs [42], or even service networking [43]. Unlike in the pure hybrid services architecture, there are no "edge boundaries" enforcing the separation between different macro-scale applications; in that respect it is a much more flat environment. However, policy can still be enforced in a distributed fashion by each individual microservice, via a Distributed Enforcement Perimeter, represented in Figure 10 with a dotted line around each microservice. Then a logical perimeter is formed around the whole set of microservices for the application.

Another technique typically used in this architecture is constraining the types of services and configurations that can be deployed. Typically, this is done through admission control mechanisms in the automation that launches jobs onto the environment. In addition, there are no hard network boundaries with limited connectivity (other than, for example, no direct connectivity to the internet).

This approach is advantageous in performance because multiple layers of costly inspection proxies can be avoided and replaced with logic in the application (e.g., in RPC client/server frameworks) in the most performance critical cases. In other cases, sidecar proxies can be deployed that are easier to provision as they scale with the workload. This approach is also a good security practice because each application is responsible for trusting other applications rather than accepting any ambient authority from being inside of the perimeter. That is, each application can make an explicit policy decision about whether to authorize a request from other applications and the policy decision can change if circumstances change (e.g., if unsigned code is deployed to a particular application, it could result in the loss of ability to authenticate to other applications). This Zero Trust security model is more robust against modern attackers [2].

Access control in such a model is achieved via enforcement within each microservice of an IAM policy based on a uniform notion of identity (e.g., mTLS certificates) provided by the compute platform. This platform-provided identity layer is a critical feature of the viability of this approach, as without it there would be no principals to define policy on (since the role of the network as a principal has been heavily attenuated in this architecture). While these primitives can be provided directly by cloud platforms, they can also be wrapped up as part of a separate product. Service mesh technologies like the open source project Istio are one example of an effort to offer these identity primitives and related service abstractions. These can be effectively combined with other technologies like application layer observability to provide visibility into the data plane traffic between services, which can help monitor compliance and/or potential violations of the defined Zero Trust policies.

An IAM policy, based on the cloud platform provided identity layer, can enforce both discretionary ACLs as well as mandatory ACLs (e.g., cloud security perimeter boundary). Note that this requires the ability to enforce policy at the application layer (or tunnel the workload through an application layer proxy). In some use cases, for example for traffic with high latency sensitivity (e.g., VoIP, gaming) or for legacy protocols (e.g., legacy UDP applications, & marketplace deployments of lift-and-shift components), the policy enforcement at application layer, based on identity, is not a realistic possibility. Such protocols can sometimes be tunneled through sidecar proxies to assert the necessary identities, but these applications still pose a challenge for the success of this Zero Trust distributed architecture. In general, the further from common HTTP/TCP based protocols one gets, the more challenging it is to use these tools and architectural approaches to achieve Zero Trust.

Additionally, a critical component in the success of this approach is ensuring the provenance of the code running on the network (since this code is itself responsible for enforcing policy). This is typically achieved by combining two controls:

1. Limit DevOps to exceptional break-glass cases for access to production, and closely audit such access.

2. Robust governance of the software supply chain and binary authorization [4]

Although the details of these controls are not the focus of the paper, it is worth noting that their combination creates a chain of trust that enables the distributed enforcement of security policy to be viable and effective. Both of these controls are challenging to pull off in practice, but when it is possible to achieve them it can create a very powerful environment for deploying highly scaled workloads. For many enterprises it is potentially more realistic to combine a Zero Trust architecture inside one of the previously considered architectures, which we will describe next.

**Section 3.4: Combined Approaches**

When enterprise applications move to the cloud, the reality is that a combination of these approaches is needed. To account for this, the approaches can be combined and nested. A few principles guide this hybrid strategy:

1. Define cloud security perimeters **abstractly**, separately from control mechanisms. Align network and identity based controls to the same definition of a cloud security perimeter.

2. Leverage **landing points** to define the outer superstructure of the cloud deployment, packaging each group of workloads within non-routable subnets.

3. Start with large groupings sufficient to migrate legacy use cases and **incrementally narrow** them.

Figure 11, shows a combined topology, where the basic super structure of Figure 4 is preserved, except that more perimeters are introduced into the picture to account for the gradation of different types of applications. Figure 11 shows on the left, a "legacy network" migrated to cloud, with two security perimeters, the Green1 and the Yellow1, which are mapped to private routable networks. The access across them is controlled by the policy enforcement provided by security gateways, as outlined in the lift-and-shift architecture in Section 3.1. On the right of the figure, there are two "modern" cloud security perimeters, the Yellow2 and the Green2, which both enclose distributed microservice architectures as outlined in the Zero Trust architecture in Section 3.3. In order to extend the security posture outside cloud, in this case, from on-premises, the topology includes a Cloud Landing Point (CLP), that enforces the data plane policies for the access from on-premises, to both the lift-and-shift legacy network security perimeters (Green1 and Yellow1), as well as the modern zero-trust cloud security perimeters (Green2 and Yellow2). The access through CLP leverages the networking services (Private Service Connect/Private Link) security concepts of Section 3.2.

Even though the lift-and-shift architectures in section 3.1 (represented in the Green1 and Yellow1 cloud security perimeters) are implemented with routable networks, they can be treated as a non-routable network in much the same way in the hybrid services architecture of Section 3.2, where individual workloads expose services to the landing point using service networking. As the underlying applications are modernized, they can be transitioned either into a different modern perimeter entirely or, eventually, the routability of the network containing these legacy workloads can be eliminated once all access to the workloads occurs through the service networking layer, allowing the transition from a legacy lift-and-shift cloud security perimeter into a modern Zero Trust distributed architecture.

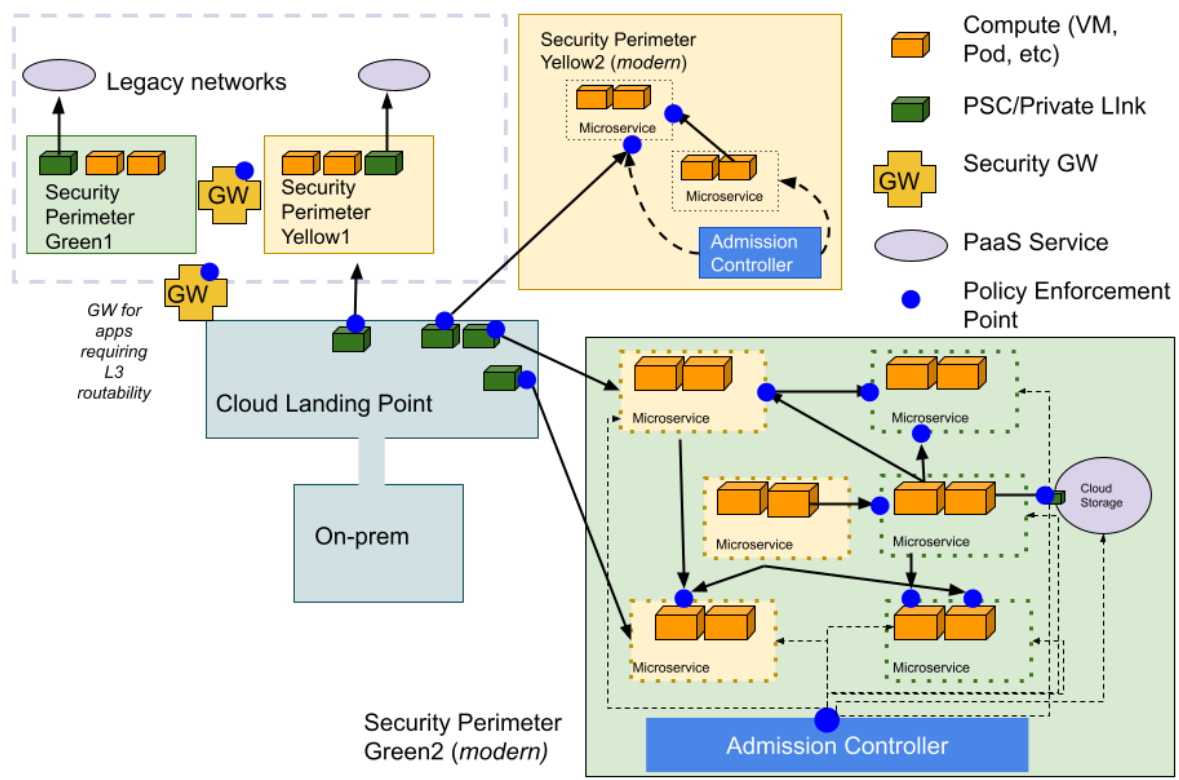

Figure 11 : Combined Architecture Example.

This example demonstrates how the concepts from all three architectures can be mixed and matched into the overall super structure enabled by the cloud security perimeter and landing point concepts. This combination provides a powerful strategy for executing a gradual cloud transition over time.

**Section 4: Comparisons and Best Practices**

We now bring all the concepts together, providing a framework that allows the security and application operators to select the architecture that achieves their cloud data plane protection goals. We start with a summary of the architectures in relation to the key concepts. We then provide a comparison of how well these architectures perform against application goals with additional insights about the use of perimeters and the mapping of the organizational concepts. We finally conclude by bringing it all together in section 4.5.

**Section 4.1: Data Plane Architectures Summary**

In Table 1, we summarize the different deployment models and how the proposed data plane architectures map to the key concepts we have identified that were described in this paper, which controls they provide, how perimeters are defined and how they use landing points.

| Data Plane Architecture | Control mechanisms | Perimeters | Landing points |
| --- | --- | --- | --- |
| **Lift-and-Shift architecture** | The controls used are typically traditional network-based. There is a limited use of application networking and identity controls. | The perimeters are based on VPC and L3/L4 firewalls. | Landing points are sometimes used as a hub to connect to hybrid locations. |
| **Hybrid services architecture** | The controls used are a combination of traditional network and cloud native ones (VPC-Service Controls, Identity Access Management). | The perimeters can be a combination of network-based perimeters (VPC, IP ranges) and cloud security perimeters based on resource hierarchy (organizations, folders, or projects). | Landing point is an inherent part of the architecture, providing a hub to extend the traditional perimeter to on-premises and provide a safe sandbox to cloud applications. Cloud native controls can be deployed **inside** a landing point for defense-in-depth. |
| **Zero Trust distributed architecture** | The controls used are typically cloud native VPC Service Controls, Identity Access Management and URL filtering. It also frequently leverages per-service RBAC authorization and other Zero Trust mechanisms. | The Zero Trust distributed architecture provides a per-service distributed enforcement, and it is often deployed within a large perimeter for defense in-depth. | Although it is possible to extend the Zero Trust architecture to access multiple clouds or on-premises, the disparity of the identity and access controls in different environments makes it more complex to manage. Therefore, Landing Point is still often used as a secure chokepoint to access the services in a given cloud from other environments. |

Table 1: How Each Architecture Utilizes the Key Concepts

In general, any large scale enterprise cloud deployment will likely include a combination of these different architectures as we outlined in Section 3.4 because each architecture serves different needs better. Some architectures (hybrid service, distributed microservice) offer better support for PaaS workloads.

Others (lift-and-shift) are often easier to integrate with an existing on-premises deployment, multiple clouds, and legacy workloads. The tools used for managing the workloads at scale also vary from highly network centric to using higher level application concepts. Table 2 explains these tradeoffs in more detail and how they differ in their ability to achieve protection and security outcomes.

| Key Data Plane Architecture Questions | How to protect from/to hybrid deployments? | How to operationally manage and audit the security posture? | How to manage organizationally? | How to apply Granular controls per workload? |
|---|---|---|---|---|
| **Related Secondary Questions** | Considerations that are unique to hybrid architecture security. | From an operational point-of-view, how does one deploy and audit at an Enterprise Scale? | How does one map cloud concepts to business concepts in terms of organizational ownership? | What level of granularity is provided for managing workload policies and access; how does this scale? |
| **Lift-and-Shift architecture** | Access between on-premises, multi-cloud and cloud is protected in this architecture through network segmentation and controls implemented by firewalls as discussed in Section 3.1. | The operation management and audit is typically done by the central network team by managing and auditing the perimeters (VPC and firewalls).<br><br>Processing of development team change requests for the central network can become an operational bottleneck. | There are one or multiple central projects, with central VPCs that are owned by cloud network and security operations teams. The developer teams are mapped to their projects that operate within each of those network VPC perimeters safely. | Typically there are no fine grained per-service controls, but instead they are per-VPC network or sometimes per subnet.<br><br>Tags can be used to apply per-service granular controls, but scalability of too fine of groups is challenging because the tags must be translated to individual IPs for enforcement in the stack. |
| **Hybrid services architecture** | Access between on-premises, multi-cloud and cloud is protected through network segmentation and controls in the cloud landing point network, often using service networking as discussed in Section 3.2. | The management and the audit includes a combination of network perimeter for the landing points and cloud security perimeters and controls within the cloud. It requires a combination of auditing perimeters and cloud native controls within the cloud. | There is one central project or account, with central VPCs, in the landing point that is owned by cloud network and security operations teams. Each of the development teams has autonomy in its own project and VPC. This model then uses Private Service Connect or Private Link in a central landing point VPC, to scale the communication of services across VPCs in the cloud. | Per-service controls can be granularly applied within the cloud using identity controls. The communication from/to hybrid environments is based on large cloud network perimeters, in the landing points. |
| **Zero Trust distributed architecture** | Access between on-premises, multi-cloud and cloud is protected through identity and application controls. This requires a strong identity root of trust. This is discussed in Section 3.3. | The management audit is usually based on cloud security perimeters and per-application identity-based access controls. The controls are typically distributed, and the audit requires scanning of the cloud deployments to surface deviations. | The protection of apps is identity and application-based, often within the Hybrid Services architecture. Frequently, many of the controls are delegated to DevSecOps teams dispersed in the organization. | Per-service controls can be granularly applied within the cloud and hybrid communication. Management is easier because the constructs are application-level rather than network-level. But centralization is more challenging because of the distributed enforcement model. |

Table 2: Key practical management aspects for each architecture

At a high level, Table 2 explains how each approach is managed in practice. It summarizes how the lift-and-shift approach is managed much the same way as a traditional enterprise network from a security and monitoring point of view, through central networking and security teams. The hybrid services architecture on the other hand enables several operating models leveraging a combination of network security controls (which can be managed centrally) and delegated management by development teams. Finally, Table 2 discusses how the Zero Trust distributed microservices model is usually easier to manage at the application layer but is more aligned towards a delegated DevSecOps model due to its distributed enforcement.

**Section 4.2: Mapping Organizational Concepts to Cloud Concepts**

As discussed in Table 1, one key variance between data plane architectures is how organizational concepts are mapped to cloud concepts. In the lift-and-shift architecture, a flat VPC network structure is typically adopted. While responsibility for this network is typically with central security/networking/cloud teams, the developer teams do have dependencies and requirements that need to be fulfilled, which can slow down developers and create bottlenecks. Division of labor is accomplished by mapping teams to projects with different IAM roles (e.g., ability to manage subnets versus the entire network) and using features like Shared VPC in GCP for multiple teams (services) to communicate with centrally managed security (e.g., firewalls) [58].

By contrast, in the hybrid services architecture, the segmentation between centrally managed components and per development team components is done using separate projects/accounts and Private Service Connect/Private Link, which is a more complete isolation boundary. As such, the governance structure can provide more flexibility while still enforcing mandatory guardrails. This scalability and ability to offer flexibility while preserving isolation is one of the main benefits of the landing point concept and the hybrid services architecture.

The Zero Trust distributed architecture is often deployed within cloud security perimeters, similar to the hybrid services architectures, but in this case, the segmentation can be more granular, typically per application (one or more microservices) instead of per development team. In other words, each application can be deployed on its own project/account, with each development team typically owning multiple projects/accounts, one per application. This can lead to tens or hundreds of projects per team, and thousands of projects in a given organization, which can be complex to manage. Technologies like Kubernetes namespaces can be used to segregate workloads within a single cluster (within a single project/account), reducing the total number of projects/accounts to manage. The Kubernetes control plane can implement admission controls to manage how multiple teams deploy their workloads onto a single cluster. Similarly, Kubernetes RBAC can be used to delegate a subset of permissions on the cluster to development teams. For GCP, there are documented best practices for this kind of Enterprise Multi-Tenancy when using Kubernetes based microservices [14].

**Section 4.3: Choosing your Perimeter**

Throughout this paper, we have focused heavily on the need to define perimeters as a logical boundary to group related workloads. However, we have not discussed in detail the factors by which one should use to decide how many perimeters to create and what workloads to assign to each one. The first step is to define a threat model of the capabilities of adversaries and base decisions on that. In our case, we are focused on sophisticated adversaries seeking unauthorized access to infrastructure data as well as insiders seeking to exfiltrate sensitive information.

Perimeters can be defined based on several different dimensions of security. Here are several factors to consider:

- Production versus non-production environments.
- Sensitivity of the data that the workload processes (e.g., crown jewels, customer data, business operations data, non-sensitive data).
- Compliance requirements (e.g., PCI, HIPAA).
- Business Unit Ownership (e.g., consumer banking versus investment banking versus risk units).
- Density of interconnection between workloads (cross-perimeter communication is more complicated than intra-perimeter communication).
- Per application, where an application consists of a set of resources and services working together to serve a single business purpose.

For example, one might create two large perimeters (dev and prod) for most data classifications and then an additional set of perimeters for high sensitivity workloads. Alternatively, one might consider creating a much larger number of very small perimeters. However, the tools used to define perimeter boundaries are often not well suited for microsegmentation or being the only line of security between workloads. Zero Trust principals also caution against being overly reliant on the perimeter as the only defense mechanism. Instead, we recommend constructing a smaller number of large perimeters using the concepts referenced here together with other intra-perimeter mechanisms to secure communication between microservices and workloads. This is especially true if one is leveraging Zero Trust mechanisms to isolate workloads from each other without relying on network micro-segmentation as the tool to achieve isolation. Exceptions can be created at perimeter boundaries for cross-perimeter communication, frequently at a higher cost both in terms of management and performance. Obviously, there is no one size fits all approach here and in some cases it can make sense to use the cloud security perimeter concepts to define small boundaries at the level of individual workloads–thus becoming a primary means of achieving isolation and reduction of trust surface area.

In summary, the perimeter is a last line of defense both at the network infrastructure layer as well as the cloud native layer, but not the only line of defense. Even in a pure Zero Trust architecture, it is also useful as a last line of defense against accidental or malicious misconfiguration. Breaching the perimeter must not imply full access to infrastructure and the ability to move laterally. Enterprises should strive, where possible, for Zero Trust security principles within a perimeter, where each component of infrastructure is able to authenticate and authorize access to it as necessary. However, the perimeter is a useful way of grouping workloads with different data sensitivity classifications and thus different policies for guardrails and controls.

**Section 4.4: Bringing it all together**

No single architectural choice will meet the security goals of all applications. The choice is dependent on where the application and organization is with the evolution to cloud native architectures, and what are the existing non-cloud hybrid environments, workloads and security models that need to work together with cloud workloads in this journey.

Google has outlined Zero Trust security as the model in its corporate and production environments, and the Zero Trust architecture (Section 3.3) is also the north star for security in the cloud. This relies on the ability to apply rich granular and contextual controls in the communication across applications in the cloud, as well as empower DevOps teams with more agility, visibility and controls. In contrast, the Zero Trust architecture does not typically extend well to other existing legacy environments, and therefore may not always be a practical choice to implement in the cloud from day one. The last consideration for the choice of model to use is the organization's roles and responsibilities. The Zero Trust architecture assumes lighter centralized teams and more empowered DevOps teams, which requires establishment of config and policy-as-code, as a way to automate compliance checks in the DevOps pipelines.

We consider that the evolution towards the Zero Trust architecture is a journey, and conclude with the following points:

- For enterprises that want to continue and resemble as much as possible their traditional security models to the cloud, the lift-and-shift architecture (section 3.1) provides a way to achieve that.

- For enterprises that are architecturally and organizationally able to adopt cloud native models, the Zero Trust architecture (section 3.3) is the recommended solution.

- For most of the cloud platform and security operators, given some level of existing hybrid architectures, we recommend starting with the hybrid services architecture (section 3.2) for new cloud deployments with its use of landing points and cloud security perimeters. Lift-and-shift (section 3.1) and Zero Trust architectures (section 3.3) can be embedded into this architecture to handle both legacy and modern workloads as described in section 3.4. The hybrid services architecture provides an opinionated framework that developers can deploy into with appropriate security guardrails to limit misconfiguration, unauthorized access, and data exfiltration risk.

**Section 5: Related Work**

Google has previously outlined approaches to Zero Trust network security in its corporate [15,16] and production environments [3]. However, that work is primarily a statement of our existing infrastructure and is not as prescriptive around adoption of these approaches in a modern public cloud. [27] conducted a recent survey of the Zero Trust literature, exploring both conceptual issues and organizational benefits as well as migration strategies. In addition, NIST has proposed a model for Zero Trust security in [2] and recently CISA announced its draft technical reference architecture for cloud security and Zero Trust in [22]. These documents discuss at a high level the goals of Zero Trust protection and catalog approaches taken by various vendors in the market. Our work proposes a mechanism for extending Zero Trust ideas into production cloud environments and incrementally evolving from a lift-and-shift network to a full identity based Zero Trust environment.

The network security literature is rich with discussions on related work. Approaches to modeling network segmentation, which is related to defense in depth and some implementations of Zero Trust, have been covered in [17,20,28] for the cloud context and by [18] in the context of a modern data center. [29] explores building a formal model of network segmentation and a mechanism for translating this model into real firewall policies in a Software Defined Network. [36] explores the more recent concept of a Software Defined Perimeter as a way of translating a logical perimeter model into concrete network segments. [43] surveys the initial shift towards network services in cloud computing. These works primarily focus on models and implementation of network segmentation in on-premises and/or IaaS environments. They do not explore the newer concepts related to PaaS services and cloud security perimeters, which we describe in our paper.

For traditional networks, firewall management is a mature field and much work covers how to manage firewalls at scale going back several decades. For example, [19] explores how to implement a declarative concept of private zones using firewall rules. Private zones are similar in concept to network perimeters, but are translated in a traditional network to IPTables rules that act on 5-tuples rather than cloud native data model concepts. [21] proposes an approach for creating conflict free firewall rules to handle the complexity of rule explosion. In terms of management, [30] explores a tool for automatic discovery of firewall rule conflicts and techniques for safely editing the ruleset while [33] proposes an analysis tool to perform custom queries. Outside of the strict context of firewall rules, there is a large volume of research into general policy scalability and conflict management, for example in the context of IPSec [31] and RBAC [32]. Our work builds on and is different from this prior work by applying this to public cloud and for both IaaS and PaaS infrastructure. This previous work focused primarily on implementing these concepts in a traditional layer 3 network.

In the identity space, [23] and [24] survey different approaches to identity management in cloud environments. They define centralized identity management based approaches and federation based approaches. Our work explores how both of these models function in a modern public cloud and can help with the implementation of security concepts like cloud security perimeters. [23] and [24] build upon a long lineage of thinking about identity. As an example, [25] surveys different theoretical models of identity systems that are still widely relevant in the cloud environment. There has also been much work on practical standards for deployments such as Kerberos (used in Microsoft Windows) [26], OICD [34], and SAML 2.0 [35], which are all widely used in enterprises as well as many universities.

Public cloud vendors like AWS [10] and Azure [11] have also published reference architectures for hub and spoke networking that share some commonality with the network design of our landing points. Both also outline appropriate ways to leverage their resource hierarchies for policy management. Our contribution is to introduce the ideas of cloud security perimeters and landing points as tools to transition from a lift-and-shift architecture to a modern, identity based Zero Trust architecture. To do this, we survey the high level set of security architecture options for cloud data planes as well as discuss the purpose of different policy management and firewall capabilities for securing these architectures, how to incorporate heterogeneous compute types, and how to transition towards a Zero Trust architecture.

One related topic we did not have the space to focus on in this paper is Access Transparency. In public cloud deployments, security operators and application owners are concerned and require visibility about the potential ability of the cloud provider to access their applications and data. This is typically addressed with solutions like Cloud Provider Access Transparency (e.g., GCP's Access Transparency [45]).

**Section 6: Summary**

In this paper, we define a conceptual model for protecting cloud applications centered around three concepts: cloud security perimeters, cloud landing points, and Zero Trust architectures. We explored different architectures using these concepts that are implemented using security controls at various layers of the infrastructure. In Section 3.1, we started with a lift-and-shift architecture that closely tracks the on-premises, infrastructure network centric approach to security. We then introduced a hybrid services architecture in Section 3.2 that illustrated the concepts of cloud security perimeters and cloud landing points for constructing cloud deployments. To round out our discussion of the data plane, we explored a modern Zero Trust distributed architecture similar to the one used by Google's production environment in Section 3.3 and discussed how it can be secured. In section 3.4, we explored how all of these architectures offer tools that can be combined together to provide a path to modernize an enterprise and transition it to the cloud while embracing Zero Trust security. The tradeoffs inherent in each approach were analyzed in the context of how they map to different layers of the infrastructure.

Overall, we hope that this paper offers insights into high level architectural strategies for protecting cloud applications as they evolve from traditional, network-centric security towards cloud native, Zero Trust security. Each of the architectures we propose are concrete steps that can be used to incrementally modernize and transition applications towards stronger more cloud-native security. The cloud era is still in its early stages, but we believe the architectures laid out in this paper are enduring and will provide a structure that can continue to evolve into the future as enterprises continue to execute their cloud transformations.

**Acknowledgements**

We would like to thank Anna Berenberg, Sam Greenfield, Geoff Voelker, John Laham, Jon McCune, Sam McVeety, Brian Rogan, Mahesh Viveganandhan, Amol Kabe, Harvey Tuch, Dave Nettleton, Sudhir Hasbe, Guru Pangal, Uday Naik, Mayank Upadhyay and Nikhil Kelshikar for providing valuable feedback on this paper and everyone working on Google Cloud and Cloud across the industry whose work informed and inspired this paper.

**References**

1. National Institute for Standards and Technology, "Framework for Improving Critical Infrastructure Cybersecurity", April 2018 - https://nvlpubs.nist.gov/nistpubs/CSWP/NIST.CSWP.04162018.pdf
2. S. Rose, O. Borchert, S. Mitchell, S. Connelly, "NIST Special Publication 800-207: Zero Trust Architecture", August 2020 - https://doi.org/10.6028/NIST.SP.800-207
3. Google Beyond Prod - https://cloud.google.com/security/beyondprod
4. Google Binary Authorization for Borg - https://cloud.google.com/security/binary-authorization-for-borg
5. A. Verma, L. Pedrosa, M. Korupolu, D. Oppenheimer, E. Tune, J. Wilkes, "Large-scale cluster management at Google with Borg", In Proceedings of the European Conference on Computer Systems (EuroSys), ACM, Bordeaux, France, 2015
6. Google VPC Service Controls - https://cloud.google.com/vpc-service-controls
7. H. Hamed and E. Al-Shaer, "Taxonomy of conflicts in network security policies," in IEEE Communications Magazine, vol. 44, no. 3, pp. 134-141, March 2006, doi: 10.1109/MCOM.2006.1607877. - https://ieeexplore.ieee.org/abstract/document/1607877
8. Azure Virtual Network Service Endpoints - https://docs.microsoft.com/en-us/azure/virtual-network/virtual-network-service-endpoints-overview
9. AWS VPC Endpoints - https://docs.aws.amazon.com/vpc/latest/privatelink/vpc-endpoints.html
10. Amazon Web Services, "Building a Scalable and Secure Multi-VPC AWS Network Infrastructure", June 2020 - https://docs.aws.amazon.com/whitepapers/latest/building-scalable-secure-multi-vpc-network-infrastructure/building-scalable-secure-multi-vpc-network-infrastructure.pdf
11. Hub-spoke network topology - https://docs.microsoft.com/en-us/azure/architecture/reference-architectures/hybrid-networking/hub-spoke?tabs=cli
12. Requested-manged network interfaces - https://docs.aws.amazon.com/AWSEC2/latest/UserGuide/requester-managed-eni.html
13. Why Container Security Matters to your Business https://services.google.com/fh/files/misc/why_container_security_matters_to_your_business.pdf
14. Best practices for enterprise multi-tenancy. https://cloud.google.com/kubernetes-engine/docs/best-practices/enterprise-multitenancy
15. R. Ward and B. Beyer, "BeyondCorp A New Approach to Enterprise Security," ;login:, vol. 39, no. 6, Dec 2014.
16. B. Osborn, J. McWilliams, B. Beyer and M. Saltonstall, "BeyondCorp; Design to Deployment at Google," ;login:, vol. 41, no. 1, 2016.

17. A. Gontarczyk, P. McMillan and C. Pavlovski, "Blueprint for Cybersecurity Zone Modelling," IT in Industry, vol. 3, no. 2, 2015.
18. Vmware, "Data Center Micro-Segmentation: A Software Defined Data Center Approach for a Zero Trust Security Strategy," VMware, 2104.
19. J. Lobo, M. Marchi and A. Provetti, "Firewall Configuration Policies for the Specification and Implementation of Private Zones," in Proceedings IEEE International Workshop on Policies for Distributed Systems and Networks, 2012.
20. R. Vanickis, P. Jacob, S. Dehghanzadeh and B. Lee, "Access Control Policy Enforcement for Zero-Trust-Networking," *2018 29th Irish Signals and Systems Conference (ISSC)*, 2018, pp. 1-6, doi: 10.1109/ISSC.2018.8585365.
21. B. Zhang, E. Al-Shaer, R. Jagadeesan, J. Riely and C. Pitcher, "Specifications of A high-level conflict-free firewall policy language for multi-domain networks," in ACM Symposium on Access Control Models and Technologies, Monterey, 2007.
22. Cybersecurity and Infrastructure Security Agency, US Digital Service, and FedRAMP. "Cloud Security Technical Reference Architecture". August 2021. https://www.cisa.gov/publication/cloud-security-technical-reference-architecture
23. N. Selvanathan, D. Jayakody, and V. Damjanovic-Behrendt. 2019. Federated Identity Management and Interoperability for Heterogeneous Cloud Platform Ecosystems. In Proceedings of the 14th International Conference on Availability, Reliability and Security (ARES '19). Association for Computing Machinery, New York, NY, USA, Article 103, 1–7.
24. B. Zwattendorfer, T. Zefferer and K. Stranacher, 2014. An Overview of Cloud Identity Management-Models. In Proceedings of the 10th International Conf. on Web Information Systems and Technologies (WEBIST), pp. 82-92
25. Y. Cao, and L. Yang, 2010. A survey of Identity Management technology. In Proceedings of the IEEE ICITIS 2010, pp. 287– 293. IEEE.
26. C. Neuman, T. Yu, S. Hartman, and K. Raeburn, 2005. The Kerberos Network Authentication Service (V5). RFC 4120 (Proposed Standard).
27. C. Buck, C. Olenberger, A. Schweizer, F. Völter, T. Eymann, Never trust, always verify: A multivocal literature review on current knowledge and research gaps of zero-trust, Computers & Security, Volume 110, 2021.
28. L. Ferretti, F. Magnanini, M. Andreolini, M. Colajanni, Survivable Zero Trust for cloud computing environments, Computers & Security, Volume 110, 2021.
29. N. Mhaskar, M. Alabbad, R. Khedri, A Formal Approach to Network Segmentation, Computers & Security, Volume 103, 2021.
30. E. S. Al-Shaer and H. H. Hamed, "Modeling and Management of Firewall Policies," in *IEEE Transactions on Network and Service Management*, vol. 1, no. 1, pp. 2-10, April 2004, doi: 10.1109/TNSM.2004.4623689.
31. Z. Fu et al., "IPSec/VPN Security Policy: Correctness, Conflict Detection, and Resolution," Proc. Policy'2001 Wksp., Jan/ 2001.
32. E. Lupu and M. Sloman, "Conflict Analysis for Management Policies," Proc. IFIP/IEEE Int'l. Symp. Integrated Network Management (IM 1997), May 1997.
33. P. Eronen and J. Zitting, "An Expert System for Analyzing Firewall Rules," Proc. 6th Nordic Wksp. Secure IT-Systems (NordSec 2001), Nov/ 2001.
34. N. Sakimural et al, OpenID Connect Core 1.0, Nov. 8, 2014. https://openid.net/specs/openid-connect-core-1_0.html
35. S. Cantor et al. Assertions and Protocols for the OASIS Security Assertion Markup Language (SAML) V2.0, OASIS Standard, Mar. 15 2005. http://docs.oasis-open.org/security/saml/v2.0/saml-core-2.0-os.pdf
36. A. Moubayed, A. Refaey and A. Shami, "Software-Defined Perimeter (SDP): State of the Art Secure Solution for Modern Networks," in *IEEE Network*, vol. 33, no. 5, pp. 226-233, Sept.-Oct. 2019, doi: 10.1109/MNET.2019.1800324
37. Google Cloud Functions - https://cloud.google.com/functions
38. Support for Active Directory over NAT - https://docs.microsoft.com/en-us/troubleshoot/windows-server/identity/support-for-active-directory-over-nat
39. GCP Workload Identity Federation - https://cloud.google.com/iam/docs/workload-identity-federation
40. AWS Security Token Service - https://docs.aws.amazon.com/STS/latest/APIReference/welcome.html

41. C. Ghali, A. Stubblefield, E. Knapp, J. Li, B. Schmidt, J. Boeuf. "Application Layer Transport Security". Google Cloud Whitepaper. https://cloud.google.com/security/encryption-in-transit/application-layer-transport-security/resources/alts-whitepaper.pdf
42. S. Convery. Network Security Architectures. Indianapolis, IN. Cisco Press. 2004.
43. Q. Duan, Y. Yan and A. V. Vasilakos, "A Survey on Service-Oriented Network Virtualization Toward Convergence of Networking and Cloud Computing," in *IEEE Transactions on Network and Service Management*, vol. 9, no. 4, pp. 373-392, December 2012, doi: 10.1109/TNSM.2012.113012.120310.
44. Layer 7 OSI Application Layer https://osi-model.com/application-layer/
45. A. Haas, A. Rossberg, D. Schuff, B.Titzer, M. Holman, D. Gohman, L. Wagner, A. Zakai, and JF Bastien. 2017. Bringing the web up to speed with WebAssembly. In Proceedings of the 38th ACM SIGPLAN Conference on Programming Language Design and Implementation (PLDI 2017). Association for Computing Machinery, New York, NY, USA, 185–200. DOI:https://doi.org/10.1145/3062341.3062363
46. Cloud Provider Access Management. Google Cloud - https://cloud.google.com/cloud-provider-access-management/docs
47. Private Service Connect, Google Cloud https://cloud.google.com/vpc/docs/private-service-connect
48. Privatelink AWS https://aws.amazon.com/privatelink/
49. OWASP ModSecurity Core Rule Set (CRS) https://github.com/coreruleset/coreruleset/tree/v3.0/master
50. Ed Coyne, Timothy R. Weil, ABAC and RBAC:Scalable, Flexible, and Auditable Access Management May-June 2013, pp. 14-16, vol. 15 DOI Bookmark: 10.1109/MITP.2013.37
51. VPC connectors for Cloud Functions https://cloud.google.com/functions/docs/networking/network-settings#route-egress-to-vpc
52. AWS Organization concepts https://docs.aws.amazon.com/organizations/latest/userguide/orgs_getting_started_concepts.html
53. Azure Organization concepts https://docs.microsoft.com/en-us/azure/devops/organizations/
54. GCP label resources https://cloud.google.com/compute/docs/labeling-resources
55. AWS tags https://docs.aws.amazon.com/ARG/latest/userguide/tagging.html
56. Identity Aware Proxy https://cloud.google.com/iap/docs/concepts-overview
57. Google Cloud Managed AD Best Practices - https://cloud.google.com/managed-microsoft-ad/docs/best-practices
58. Google Cloud Shared VPC - https://cloud.google.com/vpc/docs/shared-vpc
59. P. Siriwardena. Edge Security with an API Gateway. In: Advanced API Security. Apress, Berkeley, CA. 2020.
60. S. Ortiz, "Getting on Board the Enterprise Service Bus," in *Computer*, vol. 40, no. 4, pp. 15-17, April 2007, doi: 10.1109/MC.2007.127.
61. The state of Secrets Sprawl of GitHub https://res.cloudinary.com/da8kiytlc/image/upload/v1615208698/StateofSecretSprawlReport-2021.pdf
62. Z. Durumeric, Z. Ma, et al. "The Security Impact of HTTPS Interception". NDSS Symposium 2017. https://jhalderm.com/pub/papers/interception-ndss17.pdf
63. X.509: Information technology - Open Systems Interconnection - The Directory: Public-key and attribute certificate frameworks". ITU. Retrieved 6 November 2019.international https://www.itu.int/rec/T-REC-X.509
64. Trust Relationship Flows in Active Directory https://docs.microsoft.com/en-us/azure/active-directory-domain-services/concepts-forest-trust#trust-relationship-flows
65. Island mode configuration network for kubernetes cluster operations https://cloud.google.com/anthos/clusters/docs/on-prem/1.8/concepts/networking
66. Domain names - Implementation and specification https://datatracker.ietf.org/doc/html/rfc1035
67. Hierarchical Firewall policies https://cloud.google.com/vpc/docs/using-firewall-policies
68. Organization policy constraints https://cloud.google.com/resource-manager/docs/organization-policy/org-policy-constraints
69. IAM Permission access boundaries in AWS https://docs.aws.amazon.com/IAM/latest/UserGuide/access_policies_boundaries.html
70. Azure Active Directory Tenants https://docs.microsoft.com/en-us/microsoft-365/education/deploy/intro-azure-active-directory
71. Kubernetes Network Policies https://kubernetes.io/docs/concepts/services-networking/network-policies/

72. Kubernetes Ingress https://kubernetes.io/docs/concepts/services-networking/ingress/
73. Web technology for developers: HTTP Proxy servers and tunneling https://developer.mozilla.org/en-US/docs/Web/HTTP/Proxy_servers_and_tunneling
74. AWS VPC Sharing https://aws.amazon.com/blogs/networking-and-content-delivery/vpc-sharing-a-new-approach-to-multiple-accounts-and-vpc-management/
75. Zero Trust Networks: Building Secure Systems in Untrusted Networks by Evan Gilman, Doug Barth. Released July 2017. Publisher(s): O'Reilly Media, Inc. ISBN: 9781491962190
76. Payment Card Industry (PCI) Data Security Standard Requirements and Security Assessment Procedures Version 3.2.1, May 2018 https://www.pcisecuritystandards.org/documents/PCI_DSS_v3-2-1.pdf?
77. Kubernetes nodes https://kubernetes.io/docs/concepts/architecture/nodes/
78. Kubernetes pods https://kubernetes.io/docs/concepts/workloads/pods/
79. Kubernetes services https://kubernetes.io/docs/concepts/services-networking/service/
80. Application Layer Transport Security , Cesar Ghali, Adam Stubblefield, Ed Knapp, Jiangtao Li, Benedikt Schmidt, Julien Boeuf https://cloud.google.com/security/encryption-in-transit/application-layer-transport-security
81. VPC network networking for AWS Lambda functions https://aws.amazon.com/blogs/compute/announcing-improved-vpc-networking-for-aws-lambda-functions/
82. GCP Service Attachment (Private Service Connect) https://cloud.google.com/vpc/docs/private-service-connect#service-attachments
83. AWS Endpoint Service (Private Link) https://docs.aws.amazon.com/vpc/latest/privatelink/endpoint-service.html
84. IPv6 support in AWS Elastic Kubernetes Service (EKS) https://aws.amazon.com/blogs/containers/amazon-eks-launches-ipv6-support/